# Multi-wavelength observations of Southern Hot Molecular Cores traced by methanol masers - I. Ammonia and 24 GHz Continuum Data

S. N. Longmore[1,2⋆], M. G. Burton[1], P. J. Barnes[3,1], T. Wong[1,2,5], C. R. Purcell[1,4], J. Ott[2,6,†]
[1] *School of Physics, University of New South Wales, Kensington, NSW 2052, Sydney, Australia*
[2] *Australia Telescope National Facility, CSIRO, PO Box 76, Epping, NSW 1710, Australia*
[3] *School of Physics A28, University of Sydney, NSW 2006, Australia*
[4] *University of Manchester, Jodrell Bank Observatory, Macclesfield, Cheshire SK11 9DL, UK*
[5] *Department of Astronomy, University of Illinois, Urbana IL 61801, USA*
[6] *National Radio Astronomy Observatory, 520 Edgemont Road, Charlottesville, VA 22903, USA*
† *Jansky Fellow, Bolton Fellow*

**ABSTRACT**
We present observations of the (1,1), (2,2), (4,4) and (5,5) inversion transitions of para-ammonia ($NH_3$) and 24 GHz continuum, taken with the Australia Telescope Compact Array toward 21 southern Galactic hot molecular cores traced by 6.7 GHz methanol maser emission. We detect $NH_3$(1,1) emission toward all 21 regions and 24 GHz continuum emission toward 12 of the regions, including 6 with no reported 8 GHz continuum counterparts.

In total, we find the 21 regions contain 41 $NH_3$(1,1) cores but around half of the regions only contain a single core. We extract characteristic spectra for every core at each of the $NH_3$ transitions and present both integrated intensity maps and channel maps for each region. $NH_3$(2,2) emission was detected toward all $NH_3$(1,1) cores. $NH_3$(4,4) emission was detected in 13 of the $NH_3$(1,1) cores with $NH_3$(5,5) emission coincident with 11 of these. The $NH_3$(4,4) and (5,5) emission is always unresolved and found at the methanol maser position. Analysis of the $NH_3$(1,1) and (2,2) line ratios suggests that the cores with $NH_3$(4,4) and (5,5) emission are warmer than the remaining cores rather than simply containing more ammonia. The coincidence of the maser emission with the higher spatial resolution $NH_3$(4,4) and (5,5) emission indicates that the methanol masers are found at the warmest part of the core. In all cores detected at $NH_3$(4,4) (with the exception of G12.68-0.18 core 4), the measured linewidth increases with transition energy. The $NH_3$(1,1) spectra of several cores show an emission and absorption component slightly offset in velocity but it is unclear whether or not this is due to systematic motion of the gas. We observe large asymmetries in the $NH_3$(1,1) hyperfine line profiles and conclude this is due to non-LTE conditions arising from a number of dense, small clumps within the beam, rather than systematic motions of gas in the cores.

Assuming the 24 GHz continuum emission is optically-thin bremsstrahlung, we derive properties of the ionised gas. The rate of Lyman continuum photons required to ionise the gas of $10^{45}$-$10^{48}s^{-1}$ suggests the continuum emission is powered by stars of mass $>8M_{\odot}$. We investigate the nature of the 24 GHz continuum sources which were not detected at 8 GHz and find that these are always coincident with both ammonia and methanol maser emission. This is in contrast to those detected at both 8 and 24 GHz which are generally offset from the methanol maser emission. We investigate the possibility that these may be hyper-compact HII regions.

Finally, we separate the cores into five groups, based on their association with $NH_3$, methanol maser and continuum emission. From the different physical properties of the cores in the groups, we discuss the possibility that these groups may represent cores at different evolutionary stages of the massive star formation process.

**Key words:** stars:formation, ISM:evolution, ISM:molecules, line:profiles, masers, stars:early type

⋆ E-mail:snl@phys.unsw.edu.au

## 1 INTRODUCTION

Massive stars play a key role in shaping the local Universe. These cosmic powerhouses drive the Galactic energy cycles and shape the Galactic environment. As a major source of turbulence (through outflows and supernovae) and prodigious ultraviolet radiation, they pour energy into the interstellar medium. Yet, at the same time, their death as supernovae produce the heavy elements that act as a Galactic cooling mechanism, as well as providing the material that may form planetary systems. The degree to which massive stars influence the physical and chemical state of a galaxy depends directly on the formation rate of the massive stars within the galaxy. However, despite much effort to understand the formation process, it remains enigmatic. The reasons for this lie in the fact that it is both difficult to observe during the crucial early stages of evolution and that the theoretical problem is complex. Observationally, the problem is three-fold: massive star formation regions are much rarer and their evolution timescales are shorter than their low-mass counterparts making it difficult to catch them in the act of forming; they are found at much greater distances (so limited spatial resolution leads to confusion problems) and they are obscured by large amounts of dust for a large fraction of their main sequence lifetimes (making it impossible to view them at optical wavelengths).

Several multi-wavelength observational studies have been conducted to find sites of massive star formation prior to ultra-compact HII (UCHII) region formation (Molinari et al. 1996, 1998, 2000; Walsh et al. 1997, 1998, 1999, 2001; Fontani et al. 2005; Beltrán et al. 2006; Sridharan et al. 2002; Beuther et al. 2002; Klein et al. 2005; Hill et al. 2005, 2006). Although the specific selection criteria vary, the aim has effectively been to find sources with: sufficient bolometric luminosities to contain massive stars; far-infrared (FIR) colours suggesting the massive stars are deeply embedded; weak or no cm-continuum emission, i.e. before the massive star has ionised the surrounding material. These surveys have successfully discovered and characterised many such objects. They are generally associated with large columns of dust detected as mm and submm continuum and also molecular line emission, indicating the environment is both cold and dense.

Maser emission from the molecules of water ($H_2O$), methanol ($CH_3OH$) and the hydroxyl radical (OH) are routinely observed toward star formation regions and the specific conditions required for their existence make them a powerful additional indicator of the region's evolutionary stage. Methanol maser emission is unique in that it is has only been found toward massive star formation regions (Walsh et al. 2001; Minier et al. 2003). Masing transitions of methanol are separated into two observationally distinct classes (Batrla et al. 1987): class I, which are generally seen offset by up to a parsec from the current powering sources, and class II, which were originally thought to be found in the immediate vicinity of the (proto)stars. Modelling shows the difference between the two classes lies in the pumping mechanism which is due to collisional excitation from shocks or cloud-cloud interactions for class I transitions and radiative excitation from infrared photons in the case of class II transitions (Cragg et al. 1992). The class II transition at 6.7 GHz has proved particularly useful for studying massive star formation regions as it is both strong and readily observed at cm wavelengths.

Previous multi-wavelength surveys toward a sample of $\sim$90 regions with 6.7 GHz methanol maser emission have found the masers are generally offset from radio continuum emission but always coincident with sub-mm dust continuum, luminous enough to contain massive stars and as such may signpost stages of the massive star formation (MSF) process prior to UCHII formation (Walsh et al. 1998, 1999, 2003). In the past few years, these regions have been the focus of several studies to further characterise their physical conditions and chemical properties. Continuum observations at submm and mm wavelengths (Hill et al. 2005, 2006) revealed further "mm-only" cores toward these regions, not coincident with either IR or cm continuum emission. These "mm-only" cores may harbour protoclusters with no massive stars, or, potentially, be at an earlier stage of the formation process, prior to the onset of methanol maser emission. Subsequent single-pointing, 3mm molecular-line observations with the Mopra telescope of seven species have classified the large scale gas properties and chemical constituents of 83 regions (Purcell et al. 2006). The detection of the hot core tracer $CH_3CN$ toward most of the cores exhibiting methanol maser emission suggests that the cores must be internally heated. Many of them also exhibit asymmetric profiles indicative of infall or outflow – further evidence for star formation.

This is the first in a series of papers to build upon these results by obtaining higher resolution observations to image the individual hot molecular cores and any associated embedded stellar populations. The mm and near/mid-infrared wavelength regimes were chosen to probe the gas and (proto-) stellar content respectively. In this paper we present an interferometric study of ammonia and 24 GHz continuum toward a subset of these regions. A representative sample was chosen with varying combinations of star-formation tracers (8 GHz radio-continuum, submm & mm-continuum and/or methanol maser emission) to investigate the possible role these tracers may play in an evolutionary sequence of the MSF process. In this paper we outline the observations, present the data, obtain characteristic $NH_3$ spectra for each core and derive the ionised gas properties from the 24 GHz continuum emission. We then compare the observed core linewidths, sizes, optical depths, intensities and line profiles. In subsequent work we will derive then compare characteristic physical properties (kinetic temperatures [from large velocity gradient modelling and rotational (Boltzmann) diagrams], column densities and masses) and kinematic structures of the gas in the cores.

### 1.1 Ammonia – a detailed probe of star formation regions

As detailed in the review by Ho & Townes (1983), ammonia ($NH_3$) is an excellent probe of the physical conditions of dense molecular gas. The ammonia molecule forms a tetrahedral structure with the nitrogen sitting above the plane formed by the hydrogen atoms. The rotational energy levels are described by the two quantum numbers J (the total angular momentum) and K (the projection of this angular momentum along the molecular axis). Each energy state is denoted $NH_3$(J,K). With decay timescales of $\sim 10^9$ s, the J = K levels are termed metastable in comparison to levels with $J > K$, which decay rapidly ($10-100$ s) via $\Delta$J = 1 far-infrared transitions and are termed nonmetastable. The possible combinations of hydrogen spin states leads to two separate ammonia species: ortho-ammonia, in which all H spins are parallel (K is a multiple of 3); and para-ammonia, where the H spins are not parallel (K not a multiple of three). The structure of the molecule coupled with molecular vibration, allows the nitrogen atom to tunnel through the plane of the three hydrogen atoms. The lowest energy of these inversion transitions are readily observed at cm wavelengths ($\sim$ 24 GHz). As metastable states are normally collisionally excited, the intensity ratio of their inversion transitions provides the rotational temperature of the gas. The kinetic temperature of the gas can then be estimated by modelling the collision rates of the ammonia with the ma-jor consituent of the cores, molecular hydrogen. The large electric quadrupole moment of the nitrogen atom causes the inversion transitions to split into distinct hyperfine components, observable as five components or 'fingers'. The ratio of the components provides a measure of the optical depth, circumventing the usual difficulties associated with determining optical depths from isotopic studies – namely calibrating over different frequency bands and assuming an isotopic ratio. Finally, ammonia is a particularly good probe for high density gas as it has a high critical density ($1.9 \times 10^4 cm^{-3}$, Stahler & Palla 2005) and is more resistant to the effects of depletion than other high density tracer molecules such as CS (e.g. Bergin & Langer 1997).

| Region | RA (J2000) | DEC (J2000) | $V_{LSR}$ ($kms^{-1}$) | D (kpc) |
|---|---|---|---|---|
| G316.81-0.06 | 14:45:26.9 | -59:49:16 | -38.7 | 2.7 |
| G323.74-0.26 | 15:31:45.8 | -56:30:50 | -49.6 | 3.3 |
| G331.28-0.19 | 16:11:26.9 | -51:41:57 | -88.1 | 5.4 |
| G332.73-0.62 | 16:20:02.7 | -51:00:32 | -50.2 | 3.5 |
| G1.15-0.12 | 17:48:48.5 | -28:01:12 | -17.2 | 8.5 |
| G2.54+0.20 | 17:50:46.5 | -26:39:45 | 10.1 | 4.4 |
| G8.68-0.37 | 18:06:23.5 | -21:37:11 | 37.2 | 4.8 |
| G10.44-0.02 | 18:08:44.9 | -19:54:38 | 75.4 | 6.0 |
| G11.94-0.15 | 18:12:17.3 | -18:40:03 | 42.6 | 4.4 |
| G12.68-0.18 | 18:13:54.7 | -18:01:41 | 56.5 | 4.9 |
| G19.47+0.17 | 18:25:54.7 | -11:52:34 | 19.7 | 1.9 |
| G23.44-0.18 | 18:34:39.2 | -08:31:32 | 101.6 | 5.9 |
| G24.79+0.08 | 18:36:12.3 | -07:12:11 | 110.5 | 7.7 |
| G24.85+0.09 | 18:36:18.4 | -07:08:52 | 108.9 | 6.3 |
| G25.83-0.18 | 18:39:03.6 | -06:24:10 | 93.4 | 5.6 |
| G28.28-0.36 | 18:44:13.3 | -04:18:03 | 48.9 | 3.3 |
| G29.87-0.04 | 18:46:00.0 | -02:44:58 | 100.9 | 6.3 |
| G29.96-0.02 | 18:46:04.8 | -02:39:20 | 97.6 | 6.0 |
| G29.98-0.04 | 18:46:12.1 | -02:38:58 | 101.6 | 6.3 |
| G30.79+0.20 | 18:46:48.1 | -01:48:46 | 81.6 | 5.1 |
| G31.28+0.06 | 18:48:12.4 | -01:26:23 | 109.4 | 7.2 |

**Table 2.** Pointing centre and $V_{LSR}$ towards each region from the methanol maser position/velocity. The adopted kinematic distance to each region is taken from Purcell et al. (2006).

## 2 OBSERVATIONS AND DATA REDUCTION

The $NH_3$(1,1)/(2,2) transitions were observed simultaneously from 2004 July 27$^{th}$ – 29$^{th}$ and the $NH_3$(4,4)/(5,5) transitions were observed simultaneously from 2005 August 25$^{th}$ – 28$^{th}$ on the Australia Telescope Compact Array (ATCA)[1]. The observing setup for each run is shown in Table 1. The pointing position (centred on the methanol maser emission), $V_{LSR}$ and adopted kinematic distances (see Purcell et al. 2006) are shown in Table 2.

Compact array configurations, H168 and H214 (baselines 61-192m and 82-247m respectively), providing both East-West and North-South baselines, were used to allow for snapshot imaging[2]. Primary and characteristic synthesised beam sizes were $\sim 2'$ and $\sim 10''$ respectively. Each source was observed for $4 \times 15$ minute cuts in each transition separated over 8 hours to ensure the best possible sampling of the $uv$-plane. Figure 1 shows typical $uv$-coverage for a source at $\delta \sim -20°$ with the H168 (left) and H214 (right) array configurations. A bright (>1.5 Jy), close (<5°) phase calibrator was observed for 3 to 5 minutes before and after each cut. PKS $1253-055$ and PKS $1934-638$ were used as the bandpass and primary calibrator, respectively, in all observations.

The data were reduced using the MIRIAD (see Sault et al. 1995) package. Bad visibilities were flagged, edge channels removed and the gains/bandpass solutions from the appropriate calibrator were applied to the visibilities. The data were Fourier transformed to form image cubes with $1''$ pixels using natural weighting. Due to large system temperatures in antennae 2 and 5 for the (4,4) and (5,5) observations (caused by water in the resonant cavity of the receiver), individual visibilities were further weighted by their on-line system temperature. The images were *CLEAN*ed down to $3\sigma$ above the noise to remove sidelobes then convolved with a Gaussian beam of the synthesised beam size to produce a final *RESTOR*ed image. Continuum emission was extracted from a low-order polynomial fit to line-free channels using *uvlin* and images were made from these visibilities in the same way. Images with sufficiently large signal to noise were self-calibrated to reduce phase errors. The MIRIAD tasks *mossen* (to calculate the primary beam response) and *maths* (to divide the beam response through the *RESTOR*ed image) were used to correct for the primary beam gain response. Finally, the map intensities were converted to brightness temperature units using the conversion factor calculated from the observing frequency and synthesised beam size in the *imstat* task. From previous observations of the primary calibrator, PKS $1934-638$[3], errors in the absolute flux scale are estimated to be ∼10%. This does not effect values derived from ratios within an individual transition or between two simultaneously observed transitions but must be taken into account when comparing independent observations.

Characteristic spectra were extracted at every transition for each core at the position of the peak $NH_3$(1,1) emission. They were then baseline subtracted and fit using the *gauss* and *nh3(1,1)* methods in CLASS[4]. The *nh3(1,1)* method fits each of the hyperfine components with Gaussian profiles offset by their known, fixed frequency difference, assuming they have equal excitation temperatures and linewidths. The optical depth of the transition, $\tau^{(1,1)}_{main}$, can then be calculated from the measured brightness temperature ratio between the main-line (central hyperfine component) and one of the satellite lines through (Ho & Townes 1983, Eq. 1),

$$\frac{T^{(1,1)}_{B,main}}{T^{(1,1)}_{B,satellite}} = \frac{1-e^{-\tau^{(1,1)}_{main}}}{1-e^{-\tau^{(1,1)}_{satellite}}} = \frac{1-e^{-\tau^{(1,1)}_{main}}}{1-e^{-a\,\tau^{(1,1)}_{main}}} \qquad (1)$$

where $a$ is 0.278 and 0.221 for the inner and outer satellite lines respectively. A full treatment and in depth analysis is required to derive the gas kinetic temperature from observations of multiple $NH_3$ inversion transitions and is therefore reserved for Paper II.

The intrinsic FWHM of each line, $\Delta V_{\rm intrinsic}$, was calculated by deconvolving the instrumental component, $\Delta V_{\rm instrument}$, from the measured linewidth, $\Delta V_{\rm measured}$, through $\Delta V_{\rm intrinsic} = (\Delta V^2_{\rm measured} - \Delta V^2_{\rm instrument})^{1/2}$. The instrumental contribution was calculated as $\Delta V_{\rm instrument} = \delta\,\Delta V_{\rm chan}$, where $\Delta V_{\rm chan}$ and $\delta$ are the velocity resolution per channel, and instrumental FWHM resolution in channels, respectively, and are given in columns 8 and 9 of Table 1.

Continuum source fluxes and angular sizes were calculated

[1] Project ID C1287

[2] Due to the much larger baseline lengths (typically >4.5 km), no data from antenna 6 was used throughout the observations

[3] see http://www.narrabri.atnf.csiro.au/calibrators/

[4] http://www.iram.fr/IRAMFR/GILDAS/

| $NH_3$ Transition | Rest Freq. (GHz) | Correlator Setting | IF | BW (MHz) | Num. Chan. | Num Pols | $\Delta$V (kms$^{-1}$) | $\delta$ | Array Config | Line RMS (mJy/beam) | Cont. RMS (mJy/beam) | Beam ($''$) |
|---|---|---|---|---|---|---|---|---|---|---|---|---|
| (1,1) | 23.6945 | FULL_8_512-64 | 1 | 8 | 512 | 2 | 0.197 | 1.21 | H168 | 18 | 0.9 | 11 |
| (2,2) | 23.7226 | | 2 | 64 | 64 | 2 | 12.661 | 1.24 | | 2 | 0.3 | 11 |
| (4,4) | 24.1394 | FULL_16_256-16_256 | 1 | 16 | 256 | 2 | 0.791 | 1.21 | H214 | 9 | 0.6 | 8 |
| (5,5) | 24.5330 | | 2 | 16 | 256 | 2 | 0.764 | 1.21 | | 9 | 0.6 | 8 |

**Table 1.** Observing setup. The bandwidth (BW), number of channels and number of polarisations are listed for the two intermediate frequencies (IF) in each of the correlator settings (see the general users guide at http://www.narrabri.atnf.csiro.au/observing for more information). $\Delta$V gives the corresponding velocity resolution per channel for each IF in kms$^{-1}$. $\delta$ gives the instrumental full width at half maximum (FWHM) resolution in channels. The H168 and H214 array configurations have baselines ranging from 61 - 192m and 82 - 247m respectively. The second and third last columns show the average spectral line rms per channel and theoretical continuum brightness sensitivity (based on a one hour observation in average weather conditions using the ATCA sensitivity calculator – http://www.atnf.csiro.au). The final column gives a characteristic synthesised beam size (for a source at $\delta = -30°$).

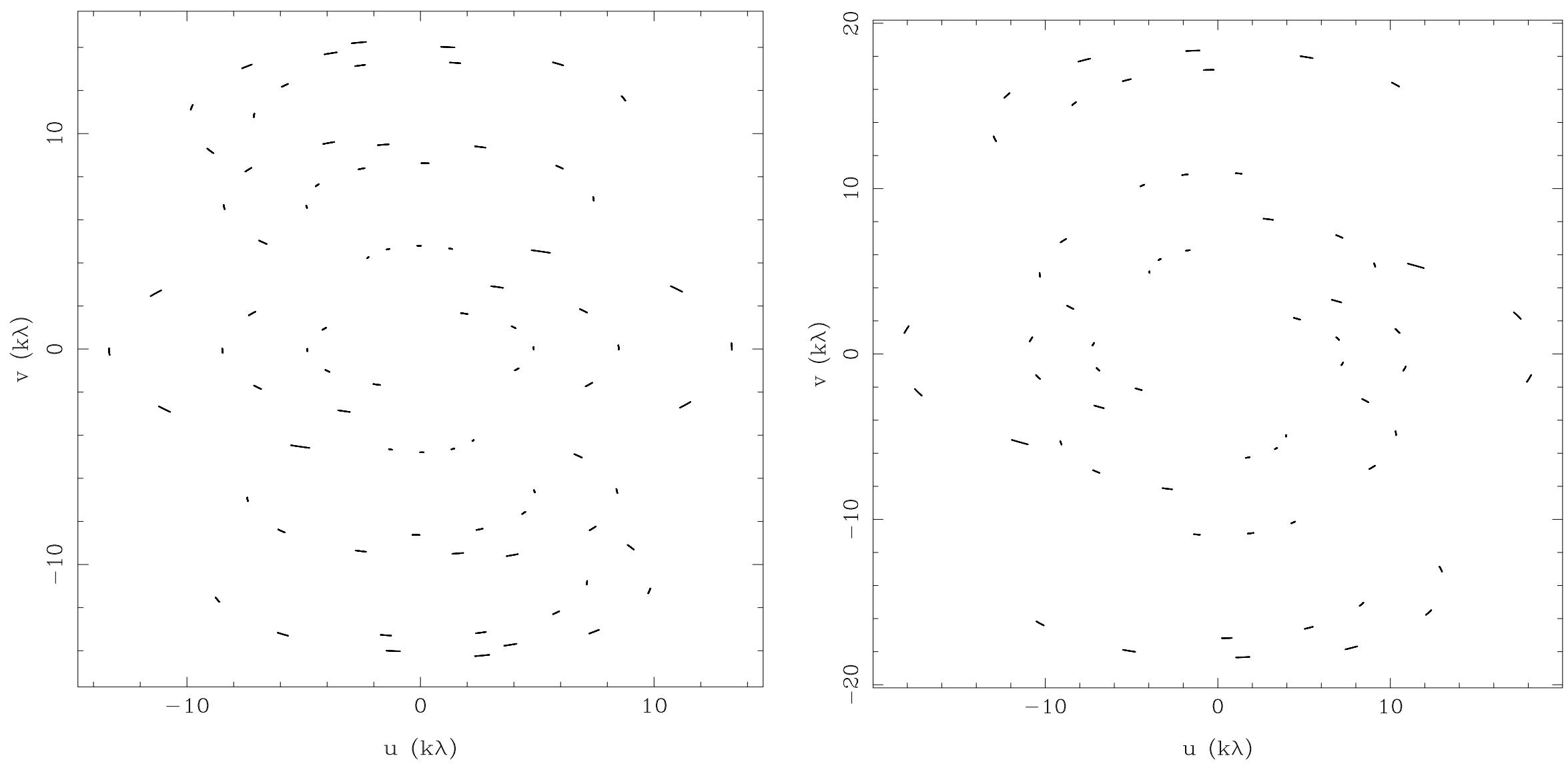

**Figure 1.** Typical $uv$-coverage for a source at $\delta \sim -20°$ with the H168 (left) and H214 (right) array configurations. Baselines that include antenna 6 have been removed.

in both the image domain and directly from the $uv$ data. The integrated flux and angular sizes of the $NH_3$ (1,1) cores from the high spectral resolution data were calculated in the image domain from the integrated intensity images. Aperture photometry was performed in the *KARMA*[5] visualisation program for sources with complicated structure. Sources with simple (point-like or Gaussian) morphologies were fit in the image and $uv$ planes using the MIRIAD tasks *imfit* and *uvfit*, which output source fluxes and deconvolved image sizes from initial estimates of the fitting parameters. Both the flux and the deconvolved source size in the image plane agreed well with those calculated from the $uv$ data. Point sources were given an upper angular size limit of 1$''$ (the minimum deconvolved angular size). The calculated continuum fluxes and angular sizes are shown in Table 3. The integrated fluxes for the low resolution $NH_3$(1,1) and $NH_3$(2,2) $\rightarrow$ $NH_3$(5,5) transitions were calculated directly from the spectra taken at the peak of the $NH_3$(1,1) emission. This removes the difficulty involved in trying to compare the integrated intensity between transitions which may have a different spatial extent.

[5] http://www.atnf.csiro.au/computing/software/karma/

## 3 DERIVING PROPERTIES FROM CONTINUUM EMISSION

At wavelengths longer than a few mm, continuum emission from massive star formation regions is expected to be dominated by free-free emission from ionised gas. Ionisation of the natal material from molecular ($H_2$) to ionised (HII ) hydrogen requires a flux of Lyman continuum (Lyc) photons ($h\nu > 13.6eV$) i.e. stars >8M$_\odot$ or earlier than B3. Following Panagia & Walmsley (1978) and Mezger & Henderson (1967), the properties of the ionised gas can be derived, assuming the continuum emission is optically thin and from a spherically symmetric, homogeneous source. The electron density ($n_e$), emission measure ($EM$) and mass of ionised gas ($M_{\rm H+}$), are then given by

$$n_e = 3.113 \times 10^2 \left(\frac{S_\nu}{Jy}\right)^{0.5} \left(\frac{T_e}{10^4 K}\right)^{0.25} \left(\frac{D}{kpc}\right)^{-0.5} \times b(\nu, T_e)^{-0.5} \theta_D^{-1.5} \mathrm{cm}^{-3} \quad (2)$$

$$EM = 5.638 \times 10^4 \left(\frac{S_\nu}{Jy}\right) \left(\frac{T_e}{10^4 K}\right) \theta_D^{-2}\, b(\nu, T_e)\ \mathrm{pc\ cm}^{-6} \quad (3)$$

| Region | Core | RA (J2000) | DEC (J2000) | $\theta_{RA}$ (arcsec) | $\theta_{DEC}$ (arcsec) | PA (∘) | $S_\nu$ (Jy) | $\langle e \rangle$ ($10^2$ cm$^{-3}$) | EM ($10^6$ pc cm$^{-6}$) | $M_{H+}$ ($M_\odot$) | $N_{Lyc}$ ($10^{47}$ s$^{-1}$) |
|---|---|---|---|---|---|---|---|---|---|---|---|
| G316.81-0.06 | 2 | 14:45:25.9 | -59:49:20.44 | 13 | 11 | -14 | 1.13 | 10.0 | 7.7 | 0.40 | 1.32 |
| G316.81-0.06 | 3 | 14:45:21.6 | -59:49:37.57 | 26 | 15 | -47 | 1.44 | 6.9 | 5.5 | 1.00 | 2.54 |
| G331.28-0.19 | 1 | 16:11:27.6 | -51:41:59.68 | 23 | 17 | -79 | 0.28 | 1.9 | 0.8 | 3.00 | 1.63 |
| G1.15-0.12 | 2 | 17:48:41.5 | -28:01:38.97 | 14 | 7 | 35 | 0.12 | 6.4 | 7.6 | 4.00 | 9.48 |
| G8.68-0.37 | 4 | 18:06:19.0 | -21:37:32.28 | 5 | 2 | 82 | 0.60 | 68.0 | 140.0 | 0.20 | 4.89 |
| G11.94-0.15 | 3 | 18:12:11.5 | -18:41:28.94 | 5 | 2 | 49 | 0.004 | 14.0 | 6.3 | 0.05 | 0.24 |
| G12.68-0.18 | 3 | 18:13:55.3 | -18:01:36.22 | 13 | 9 | 27 | 0.14 | 3.1 | 1.2 | 0.50 | 0.57 |
| G19.47+0.17 | 2 | 18:25:54.2 | -11:52:22.28 | 7 | 4 | -12 | 0.03 | 6.8 | 1.0 | 0.01 | 0.02 |
| G24.79+0.08 | 1 | 18:36:12.6 | -7:12:12.01 | 1 | 1 | 1 | 0.11 | 80.0 | 110.0 | 0.04 | 1.05 |
| G24.79+0.08 | 3 | 18:36:10.7 | -7:11:23.33 | 13 | 8 | -10 | 0.09 | 2.9 | 1.4 | 1.00 | 1.41 |
| G24.85+0.09 | 1 | 18:36:18.4 | -7:08:53.30 | 20 | 9 | -61 | 0.10 | 1.3 | 0.3 | 0.80 | 0.35 |
| G28.28-0.36 | 6 | 18:44:15.0 | -4:17:55.96 | 3 | 2 | -38 | 0.44 | 66.0 | 78.0 | 0.04 | 0.89 |
| G29.96-0.02 | 1 | 18:46:04.0 | -2:39:21.84 | 8 | 7 | -52 | 3.21 | 25.0 | 58.0 | 2.00 | 19.00 |
| G31.28+0.06 | 2 | 18:48:11.8 | -1:26:30.87 | 10 | 9 | -87 | 0.28 | 4.3 | 2.9 | 2.00 | 2.40 |

**Table 3.** Positions, measured and derived properties of observed continuum sources. $\theta_{RA}$ and $\theta_{DEC}$ give the deconvolved angular size in RA and DEC in arcsec, PA is the position angle in degrees and $S_\nu$ is the flux density calculated as outlined in § 2. $\langle e \rangle$, EM, M $_{H+}$ and $N_{Lyc}$ are the derived electron density, emission measure, ionised mass and flux of lyman continuum photons required to ionise the region (§ 3). The derivations assume the continuum emission has come from a spherical, optically thin, thermal source with electron temperature, $T_e = 10^4 K$ and an abundance ratio by number of $He^+$ to $H^+$ of 0.085 (Rubin et al. 1998). The number of Lyman continuum photons emitted per second ($N_{Lyc}$) assumes a power law approximation for the recombination coefficient of hydrogen ($\alpha_B$) such that $\alpha_B \sim 2.59 \times 10^{-13}$ $(T_e/10^4 K)^{-0.83}$ cm$^3$s$^{-1}$.

$$M_{\rm H+} = 0.7934 \left(\frac{S_\nu}{Jy}\right)^{0.5} \left(\frac{T_e}{10^4 K}\right)^{0.25} \left(\frac{D}{kpc}\right)^{2.5} \theta_D^{1.5} \times b(\nu, T_e)^{-0.5} \frac{1}{1+y} M_\odot \quad (4)$$

where $S_\nu$ is the flux density at frequency $\nu$, $T_e$ is the electron temperature of the ionised gas, D is the distance to the region, $\theta_D$ is the angular diameter of the source in arcmin and $y$ is the abundance ratio by number of $He^+$ to $H^+$. We adopt a value of $y$=0.085, as found in the Orion Nebula (Rubin et al. 1998). The function $b(\nu, T_e)$ is given by,

$$b(\nu, T_e) = 1 + 0.3195 \ \log\left(\frac{T_e}{10^4 K}\right) - 0.2130 \ \log\left(\frac{\nu}{GHz}\right) \quad (5)$$

Assuming no UV photons have leaked out of the HII region or been absorbed by dust, a lower limit to the rate of Lyman continuum photons required to keep the gas ionised can be estimated. Comparison of this lower limit with models of early-type stars (e.g. Panagia 1973) provides an approximate spectral type of the powering source(s).

## 4 RESULTS

$NH_3$(1,1) emission is detected toward all 21 regions and 24 GHz continuum emission detected towards 12 of them. Continuum images are shown in Figure 12 and discussion of their corresponding positions and observed/derived properties is given in §4.2. Figure 13 shows both the $NH_3$(1,1) channel maps of the main hyperfine satellite and $NH_3$(1,1) integrated intensity images for all the regions. The synthesised beam is shown in the bottom left corner of both images and the velocity of each plane of the channel map is shown in the top left corner. Additionally, the circle and line on the integrated intensity and continuum maps gives the primary beam size and linear scale respectively. The peak flux density is shown in both the integrated intensity and continuum images. The 24 GHz continuum emission in Figure 12 is overlayed on the $NH_3$(1,1) integrated intensity images as a dashed contour. The bold, dashed contours in G8.68-0.37 and G12.68-0.18 highlight regions of $NH_3$(1,1) absorption.

Detections within each region were separated into individual cores based on the following criteria: the core centres were first determined through Gaussian fits to the images; ammonia cores were considered distinct if separated by more than a synthesised beam width spatially, or more than the full width at half maximum (FWHM) of the narrower component in velocity if at the same sky position. Similarly, ammonia and continuum emission were considered to belong to the same core if the peaks were spatially separated by less than a synthesised beam width. Finally, cores were considered associated with 6.7 GHz methanol maser emission if they were both within a synthesised beam width spatially and the FHWM in velocity space. Based on these criteria, we find 41 $NH_3$(1,1) cores (3 of which are in absorption), with two separated in velocity (G316.81-0.06 Core 2(a) & (b)) and 14 cores of 24 GHz continuum. The cores in each region were then labelled in order of their $NH_3$(1,1) intensity as shown in Figures 12 & 13. Cores at the same spatial position but separated in velocity were additionally labelled a, b, c etc. In nearly all cases these criteria were sufficient to both unambiguously separate cores and determine their association with continuum and maser emission. The few cases where this is not so clear are discussed in §4.4 and summarised in §4.3.1. As G316.81-0.06 Core 2(b) is only seen in the high resolution $NH_3$(1,1) spectrum, it is not included in further analysis.

Figure 2 plots a histogram of the number of $NH_3$(1,1) cores found towards each region and shows that nearly half of the regions contain a single core. There are significantly fewer regions with a high multiplicity of cores. However, care was taken not to over-interpret the morphology and core multiplicity as the measured source structure of interferometric observations is dependent on the $uv$-coverage. For example, an extended source observed in a compact array configuration may be resolved into multiple separate sources or may not even be detected at all, if observed with a more extended array configuration. A limit to the largest angular extent of an object that can be imaged is given by the shortest antenna baseline pair. Using the shortest physical distance for the H168 and H214 configurations of 61/82m this corresponds to a maximum angular source size of 43 and 32′′. However, this is in fact a lower limit, as the smallest *projected* baseline on the sky will be less than the physical distance between the antennae.

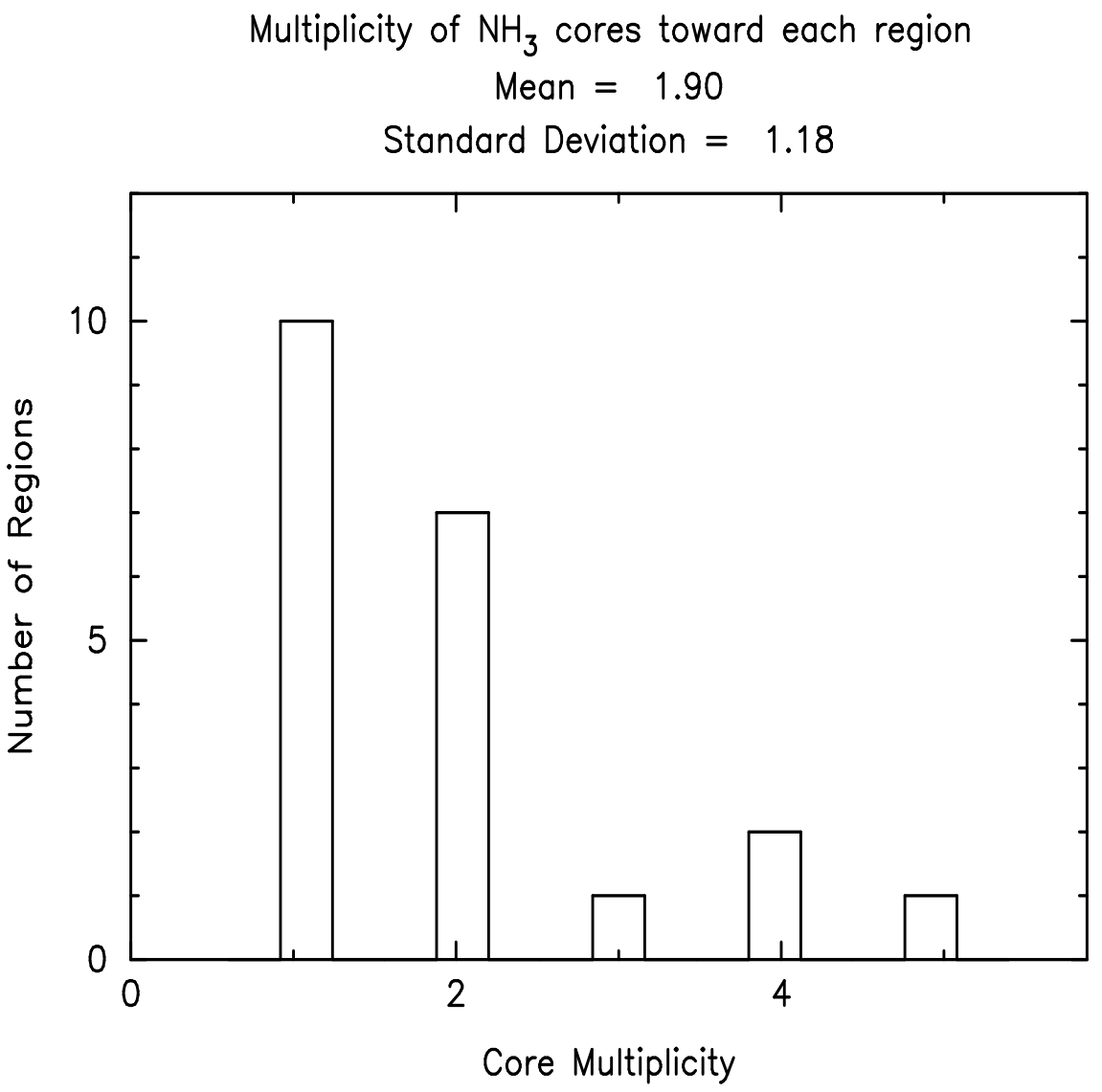

**Figure 2.** Histogram of the number of individual $NH_3$(1,1) cores (defined in §4) detected toward each region.

The morphology of the $NH_3$(2,2) emission appears very similar to that of the $NH_3$(1,1) emission, which is not unexpected due to the identical $uv$-coverage and the assumption that they come from the same gas. However, with a much lower spectral resolution ($\sim$13 kms$^{-1}$) in the $NH_3$(2,2) observations, it is not possible to compare their velocity structure. $NH_3$(4,4) emission was detected toward 13 regions with $NH_3$(5,5) emission detected at the same position in 11 of these. Imaging shows that all of these detections are from a single, unresolved source, although the lack of extended structure is due, at least in part, to using a more extended antennae configuration than the $NH_3$(1,1) and (2,2) observations with fewer short baselines. All of the $NH_3$(4,4) and (5,5) detections were found coincident (within a synthesised beam width) to both the peak $NH_3$(1,1) emission from a previously determined core and to the 6.7 GHz maser position.

## 4.1 $NH_3$ Spectra

Characteristic $NH_3$ spectra were extracted for each detected transition at the position corresponding to the peak of the $NH_3$(1,1) emission. The characteristic spectra are shown in Figure 14 and are overlayed with the profile fit as outlined in §2. With the exception of the $NH_3$(2,2) line toward G31.28+0.06 and cores with $NH_3$(1,1) hyperfine asymmetries (see §5.4) the profiles provide a good fit to the spectra. In Figure 14, spectra with low signal to noise have been Hanning smoothed. The core positions, associations and output values from the high spectral resolution $NH_3$(1,1) profile fits are given in Table 4. The output values from fits to the low velocity resolution $NH_3$(1,1) spectra and the higher transition energy spectra are given in Table 5. The apparent discrepancy between the integrated intensities of the $NH_3$(1,1) high and low spectral resolution data is due to the fact that the value in Table 4 has been integrated over the spatial extent of the core (see §2 for more details).

As shown in Figure 14, four cores (G323.74-0.26 core 1, G19.47-0.17 cores 1 & 2 and G29.96-0.02 core 1) appear to have multiple velocity components visible in the main line of the $NH_3$(1,1) spectra but are closer together in velocity than their FWHM. As such, they have previously been included as a single source under the criteria outlined in §4. Although treating each velocity component as an individual core slightly alters the number of cores in the various groups in Figure 2 and outlined in §4.3, it does not significantly affect their general distribution. However, care is taken to discard these cores in further discussion regarding core linewidths.

The lower spectral resolution $NH_3$(1,1) and (2,2) observations are denoted 1L and 2L, respectively, in Table 5. Despite the large difference in spectral resolution [12.7 kms$^{-1}$ for 1L and 2L compared to 0.8 kms$^{-1}$ for the $NH_3$(4,4) and (5,5) observations], the measured $V_{LSR}$ is similar for most cores. Inspection of Figure 14 shows that the cases where this does not hold (such as G2.54+0.20 core 1, G8.68-0.37 core 3 and G28.28-0.36 core 1) are likely to be due to low signal to noise spectra.

Several of the $NH_3$(4,4) spectra (G25.83-0.18 core 1, G29.96-0.02 core 1, G24.79+0.08 core 1, G19.47-0.17 core 1 and G12.68-0.18 core 4) show potential hyperfine structure. The nature of this emission will be discussed when deriving the physical properties of the gas in Paper II.

Comparison of the $NH_3$ spectra toward G12.68-0.18 core 4 produces some unexpected results. The results of the Gaussian profile fitting in Table 5 of the low resolution $NH_3$(1,1) and (2,2) spectra show that both the peak and the integrated $NH_3$(2,2) intensity is greater than that of the (1,1). The value and formal error to the fit of the integrated intensities (12.9$\pm$1.0 and 21.0$\pm$1.3 Kkms$^{-1}$ for $NH_3$(1,1) and (2,2) respectively) shows this is significant. The reasonable signal to noise of the spectra and reasonable image quality would also suggest this is real. Even taking into account the ratio of statistical weights of the $NH_3$(1,1) to (2,2) of 3:5, it is difficult to reconcile the intensities with the population levels of a Boltzmann distribution. Not only must the core be extremely hot, but there may also be optical depth effects involved. Alternatively, it is possible that some very cold gas lies along the line of sight which preferentially absorbs the $NH_3$(1,1) emission. However, this seems improbable as the absorption did not show up in the high spectral resolution $NH_3$(1,1) observations. The higher spectral resolution observations show the core linewidths are among the smallest of all those observed within each respective transitions ($\Delta$V = 0.8$\pm$0.07, 5.1$\pm$1.1 and 2.164$\pm$0.4 kms$^{-1}$ for $NH_3$(1,1), (4,4) and (5,5) respectively). The narrow linewidths suggests this extremely warm emission is also from a very small parcel of gas. In addition, the high resolution $NH_3$(1,1), (4,4) and (5,5) spectra show a spike of emission in a single channel. A similar spike of emission observed in the $NH_3$(5,5) transition towards G9.62+0.19 has been shown to be masing (Hofner et al. 1994). Although no corresponding maser emission was seen in $NH_3$(1,1), (2,2) or (4,4) towards this source, Wilson & Schilke (1993) predict that if masing emission is observed in one metastable state, other metastable transitions should also be masing. Higher spatial and spectral resolution observations to derive brightness temperatures are required to unambiguously determine if the emission from G12.68-0.18 core 4 is masing.

It is worth noting that the spectra from G19.47-0.17 core 1, G8.68-0.37 core 1 and G331.28-0.19 core 1 also exhibit spikey emission in the $NH_3$(5,5) transition. Again, without higher resolution observations it is not possible to determine if this emission is masing.

| Region | Core | RA (J2000) | DEC (J2000) | $T^*_B$ (K) | $V_{LSR}$ (km/s) | $\Delta V$ (km/s) | $\tau^{(1,1)}_{main}$ | $\int T^*_B dV$ (Kkm/s) | $\theta_\alpha$ ($''$) | $\theta_\delta$ ($''$) | PA ($\circ$) | Size (pc) |
|---|---|---|---|---|---|---|---|---|---|---|---|---|
| G316.81-0.06 | $1^{e}$ | 14:45:29.3 | -59:49:08.0 | 33.5±0.7 | -38.43±0.01 | 1.21±0.01 | 2.8±0.1 | 201 | 14 | 13 | -21 | 0.18 |
| G316.81-0.06 | $2^{e,a,m,c}$ | 14:45:26.6 | -59:49:16.4 | 14.2±0.4 | -38.60±0.01 | 1.89±0.03 | 2.5±0.1 | 139 | 15 | 12 | 53 | 0.18 |
| G323.74-0.26 | $1^{e,m}$ | 15:31:45.7 | -56:30:51.5 | 8.4±0.2 | -50.52±0.03 | 3.49±0.07 | 1.1±0.1 | 128 | 19 | 4 | 0 | 0.14 |
| G331.28-0.19 | $1^{e,m,c}$ | 16:11:26.7 | -51:41:54.6 | 7.0±0.3 | -87.16±0.04 | 3.76±0.10 | 1.0±0.1 | 117 | 21 | 9 | 58 | 0.37 |
| G332.73-0.62 | $1^{e,m}$ | 16:20:03.2 | -51:00:28.9 | 24.4±0.7 | -49.51±0.01 | 1.20±0.02 | 2.4±0.1 | 99 | 17 | 9 | 57 | 0.21 |
| G332.73-0.62 | $2^{e}$ | 16:20:05.7 | -51:00:06.1 | 19.8±1.1 | -49.48±0.01 | 1.02±0.03 | 4.1±0.3 | 83 | 20 | 13 | 2 | 0.27 |
| G1.15-0.12 | $1^{e,m}$ | 17:48:49.0 | -28:01:09.4 | 12.9±0.9 | -15.57±0.03 | 1.39±0.06 | 3.0±0.3 | 21 | 14 | 11 | -57 | 0.51 |
| G2.54+0.20 | $1^{e,m}$ | 17:50:47.0 | -26:39:41.7 | 17.2±1.6 | 11.51±0.02 | 0.98±0.05 | 5.0±0.7 | 47 | 12 | 8 | -89 | 0.21 |
| G2.54+0.20 | $2^{e}$ | 17:50:45.2 | -26:39:45.0 | 10.7±0.5 | 10.85±0.03 | 2.55±0.06 | 2.4±0.2 | 68 | 11 | 7 | -63 | 0.19 |
| G2.54+0.20 | $3^{e}$ | 17:50:44.4 | -26:39:57.3 | 20.6±1.1 | 10.34±0.01 | 0.87±0.03 | 3.2±0.3 | 81 | 16 | 15 | 13 | 0.33 |
| G8.68-0.37 | $1^{e,m}$ | 18:06:23.5 | -21:37:08.1 | 16.1±0.9 | 39.18±0.03 | 2.21±0.07 | 2.6±0.3 | 274 | 19 | 17 | -22 | 0.42 |
| G8.68-0.37 | $2^{e}$ | 18:06:19.4 | -21:37:19.1 | 57.0±11.3 | 37.35±0.04 | 1.32±0.08 | 18.3±4.0 | 135 | 13 | 11 | -33 | 0.27 |
| G8.68-0.37 | $3^{e}$ | 18:06:19.6 | -21:37:39.4 | 11.6±0.7 | 34.63±0.02 | 1.51±0.05 | 3.7±0.3 | 33 | 10 | 5 | 90 | 0.16 |
| G8.68-0.37 | $4^{a,m,c}$ | 18:06:19.1 | -21:37:31.1 | -14.7±0.6 | 37.72±0.03 | 2.70±0.05 | 3.3±0.2 | -44 | 8 | 7 | 87 | 0.17 |
| G10.44-0.02 | $1^{e,m}$ | 18:08:44.9 | -19:54:39.1 | 16.9±0.9 | 77.22±0.03 | 2.63±0.07 | 4.9±0.4 | 114 | 15 | 9 | -8 | 0.34 |
| G10.44-0.02 | $2^{e}$ | 18:08:45.0 | -19:54:17.1 | 7.2±0.7 | 74.98±0.05 | 1.95±0.11 | 3.3±0.5 | 33 | 11 | 10 | -35 | 0.30 |
| G11.94-0.15 | $1^{e,m}$ | 18:12:17.1 | -18:40:04.3 | 14.3±0.6 | 42.86±0.02 | 1.93±0.05 | 2.2±0.2 | 87 | 13 | 11 | 19 | 0.25 |
| G11.94-0.15 | $2^{e}$ | 18:12:19.6 | -18:39:50.2 | 9.2±0.4 | 44.92±0.02 | 1.87±0.05 | 1.8±0.2 | 52 | 10 | 8 | -24 | 0.19 |
| G12.68-0.18 | $1^{e}$ | 18:13:53.0 | -18:01:57.6 | 40.3±1.7 | 56.23±0.01 | 1.71±0.02 | 8.5±0.5 | 132 | 13 | 9 | -14 | 0.25 |
| G12.68-0.18 | $2^{e}$ | 18:13:55.7 | -18:01:21.0 | 41.9±2.5 | 57.60±0.01 | 0.95±0.02 | 7.6±0.6 | 50 | 11 | 10 | 40 | 0.25 |
| G12.68-0.18 | $3^{a,c}$ | 18:13:55.8 | -18:01:33.3 | -8.9±0.6 | 57.02±0.03 | 1.59±0.06 | 3.3±0.3 | -34 | 10 | 10 | 33 | 0.24 |
| G12.68-0.18 | $4^{e,m}$ | 18:13:54.6 | -18:01:49.3 | 6.5±1.0 | 56.07±0.04 | 0.80±0.07 | 4.3±1.0 | 22 | 5 | 5 | -61 | 0.12 |
| G19.47+0.17 | $1^{e,m}$ | 18:25:54.6 | -11:52:32.5 | 7.2±0.2 | 20.79±0.06 | 7.05±0.11 | 1.7±0.1 | 174 | 13 | 11 | -14 | 0.11 |
| G19.47+0.17 | $2^{e,c}$ | 18:25:54.4 | -11:52:19.4 | 3.6±0.3 | 18.39±0.07 | 3.17±0.14 | 2.3±0.4 | 24 | 11 | 9 | 87 | 0.09 |
| G23.44-0.18 | $1^{e,m}$ | 18:34:39.1 | -8:31:36.3 | 24.4±0.5 | 102.30±0.01 | 2.67±0.03 | 2.8±0.1 | 385 | 18 | 13 | 8 | 0.44 |
| G24.79+0.08 | $1^{e,m,c}$ | 18:36:12.5 | -7:12:08.9 | 36.6±0.6 | 111.27±0.02 | 4.59±0.04 | 3.4±0.1 | 564 | 16 | 12 | 24 | 0.52 |
| G24.79+0.08 | $2^{e}$ | 18:36:13.0 | -7:12:03.8 | 20.2±0.7 | 110.57±0.03 | 2.55±0.06 | 2.3±0.1 | 134 | 24 | 10 | -39 | 0.58 |
| G24.85+0.09 | $1^{e,m,c}$ | 18:36:18.6 | -7:08:50.4 | 24.1±0.6 | 110.02±0.01 | 1.26±0.02 | 3.6±0.1 | 46 | 6 | 3 | 63 | 0.13 |
| G24.85+0.09 | $2^{e}$ | 18:36:19.7 | -7:08:54.5 | 1.7±0.4 | 110.73±0.06 | 1.53±0.22 | 0.3±0.6 | 22 | 16 | 14 | 67 | 0.46 |
| G25.83-0.18 | $1^{e,m}$ | 18:39:03.4 | -6:24:10.6 | 17.9±0.3 | 94.77±0.02 | 4.20±0.04 | 2.4±0.1 | 235 | 11 | 9 | -64 | 0.28 |
| G28.28-0.36 | $1^{e,m}$ | 18:44:13.8 | -4:18:07.9 | 40.1±1.4 | 50.61±0.01 | 0.66±0.01 | 4.9±0.2 | 29 | 16 | 8 | 62 | 0.18 |
| G28.28-0.36 | $2^{e}$ | 18:44:12.6 | -4:18:01.8 | 29.7±3.2 | 50.98±0.02 | 0.77±0.04 | 9.4±1.3 | 77 | 18 | 15 | -55 | 0.26 |
| G28.28-0.36 | $3^{e}$ | 18:44:11.3 | -4:17:13.3 | 23.3±1.0 | 49.80±0.01 | 1.03±0.03 | 4.6±0.3 | 65 | 15 | 12 | -29 | 0.21 |
| G28.28-0.36 | $4^{e}$ | 18:44:11.6 | -4:17:00.2 | 42.5±3.8 | 50.05±0.01 | 0.92±0.03 | 13.7±1.4 | 43 | 13 | 10 | -20 | 0.18 |
| G28.28-0.36 | $5^{e}$ | 18:44:16.6 | -4:19:04.8 | 14.0±1.3 | 49.29±0.02 | 0.66±0.04 | 5.1±0.7 | 33 | 16 | 13 | -85 | 0.23 |
| G29.87-0.04 | $1^{e,m}$ | 18:45:59.4 | -2:45:07.5 | 4.5±0.4 | 103.09±0.05 | 2.27±0.12 | 0.9±0.3 | 29 | 17 | 4 | 81 | 0.26 |
| G29.96-0.02 | $1^{e,m,c}$ | 18:46:03.7 | -2:39:21.8 | 17.1±0.4 | 98.27±0.03 | 3.76±0.05 | 2.2±0.1 | 174 | 11 | 7 | 48 | 0.25 |
| G29.98-0.04 | $1^{e,m}$ | 18:46:12.8 | -2:39:02.3 | 23.8±0.7 | 102.64±0.02 | 2.06±0.04 | 3.1±0.1 | 273 | 26 | 11 | 6 | 0.52 |
| G30.79+0.20 | $1^{e,m}$ | 18:46:47.8 | -1:48:58.8 | 26.9±1.0 | 82.81±0.02 | 2.12±0.04 | 4.3±0.2 | 235 | 17 | 12 | 33 | 0.36 |
| G31.28+0.06 | $1^{e,m}$ | 18:48:13.7 | -1:26:17.3 | 15.1±0.5 | 111.38±0.02 | 2.42±0.05 | 1.4±0.1 | 42 | 18 | 9 | -35 | 0.45 |

**Table 4.** Observed positions and properties of the cores detected in the high spectral resolution $NH_3$(1,1) observations. The superscript e, a, m and c in the second column denote $NH_3$ emission, $NH_3$ absorption, association with methanol maser emission and association with continuum emission, respectively. $T^*_B$ is the peak brightness temperature, $V_{LSR}$ is the velocity of the central quadrupole component, $\Delta V$ is the deconvolved line width (see §2) and $\tau^{(1,1)}_{main}$ is the main line optical depth derived from the CLASS program (as outlined in § 2). $\int T^*_B dV$ gives the brightness temperature integrated over the angular extent of the core in the integrated intensity images, as outlined in § 2. $\theta_{RA}$ and $\theta_{DEC}$ are the deconvolved angular extent of the source in RA and DEC measured in arcsec and PA is the position angle in degrees. The final column gives the corresponding core size assuming spherical symmetry [angular size of $(\theta_\alpha \theta_\delta)^{1/2}$] and given the distance to the region shown in Table 2. An error in the core size of 10% is inherent from the uncertainty in the distance.

## 4.2 Continuum Emission

The 24 GHz continuum detections are shown in Figure 12 with the position of the 6.7 GHz methanol maser emission and 8 GHz continuum detections (Walsh et al. 1998) illustrated as crosses and boxes respectively. The observed properties (position, size and flux) and derived properties (outlined in §3) are shown in Table 3. Several sources (G1.15-0.12 core 2, G11.94 core 3 and G8.68-0.37 core 4) lie outside the half-power beam width (HPBW) of the primary beam and therefore have larger uncertainties in the measured fluxes. In general, the emission is seen to come from a few centrally bright cores with simple morphology. However, a full discussion of continuum emission morphology requires higher image fidelity than the limited $uv$-coverage obtained in snapshot imaging mode.

Of the 14 continuum cores detected at 24 GHz, 10 are within 2 synthesised beams of the 6.7 GHz methanol maser emission. This is contrary to the results of Walsh et al. (1998), who found the masers generally offset from 8 GHz continuum emission. However, six of the 24 GHz continuum sources at the site of the methanol maser emission have no 8 GHz counterparts while the majority of cores not associated with methanol masers are detected at 8 GHz. This suggests there may be a difference in the population of cores detected at only 24 GHz to those detected at both 8 and 24 GHz.

For all the 24 GHz continuum sources, the electron densities, $n_e$, and emission measures, EM, of $\sim 10^3 cm^{-3}$ and $\sim 10^7$ pc $cm^{-6}$

| Region | Core | Transition | $\int T_B^* dV$ (K.km/s) | $V_{LSR}$ (km/s) | $\Delta V$ (km/s) | Peak (K) |
|---|---|---|---|---|---|---|
| G316.81-0.06 | 1 | 1L | 133.5±3.0 | -37.3±0.3 | 28.7±0.8 | 4.37 |
| G316.81-0.06 | 1 | 2L | 37.3±2.4 | -38.5±0.7 | 16.2±1.4 | 2.17 |
| G316.81-0.06 | 2 | 1L | 38.5±2.7 | -38.1±1.2 | 38.2±3.0 | 0.95 |
| G316.81-0.06 | 2 | 2L | 23.1±1.7 | -42.2±0.8 | 16.7±1.2 | 1.30 |
| G316.81-0.06 | 2 | 4 | 7.5±0.8 | -40.0±0.3 | 5.3±0.7 | 1.30 |
| G316.81-0.06 | 2 | 5 | 4.7±0.7 | -40.3±0.4 | 5.3±1.1 | 1.30 |
| G323.74-0.26 | 1 | 1L | 70.7±2.3 | -49.1±0.4 | 23.9±1.0 | 2.78 |
| G323.74-0.26 | 1 | 2L | 32.8±2.0 | -51.2±0.6 | 17.2±1.0 | 1.80 |
| G323.74-0.26 | 1 | 4 | 5.9±0.4 | -51.2±0.2 | 4.7±0.3 | 1.15 |
| G323.74-0.26 | 1 | 5 | 9.0±0.6 | -51.7±0.2 | 7.3±0.6 | 1.16 |
| G331.28-0.19 | 1 | 1L | 59.9±1.2 | -86.7±0.2 | 25.9±0.7 | 2.18 |
| G331.28-0.19 | 1 | 2L | 43.6±1.0 | -88.1±0.2 | 19.7±0.5 | 2.08 |
| G331.28-0.19 | 1 | 4 | 13.0±0.8 | -87.3±0.2 | 8.1±0.6 | 1.50 |
| G331.28-0.19 | 1 | 5 | 4.4±0.3 | -87.5±0.1 | 4.2±0.5 | 0.98 |
| G332.73-0.62 | 1 | 1L | 36.6±1.5 | -49.7±0.5 | 24.1±1.2 | 1.43 |
| G332.73-0.62 | 1 | 2L | 14.2±1.3 | -52.5±1.1 | 15.8±2.8 | 0.84 |
| G332.73-0.62 | 2 | 1L | 26.0±0.8 | -49.4±0.6 | 30.8±1.1 | 0.79 |
| G332.73-0.62 | 2 | 2L | 6.1±0.5 | -52.4±0.6 | 13.1±0.4 | 0.43 |
| G1.15-0.12 | 1 | 1L | 29.8±1.1 | -15.8±0.5 | 28.2±1.2 | 0.99 |
| G1.15-0.12 | 1 | 2L | 11.8±1.1 | -18.1±1.0 | 23.3±3.1 | 0.47 |
| G2.54+0.20 | 1 | 1L | 1.6±0.2 | 2.5±1.1 | 17.8±2.2 | 0.09 |
| G2.54+0.20 | 1 | 2L | 1.1±0.1 | 13.5±1.3 | 12.7±0.5 | 0.08 |
| G2.54+0.20 | 2 | 1L | 12.7±0.2 | 10.3±0.2 | 22.5±0.3 | 0.53 |
| G2.54+0.20 | 2 | 2L | 1.9±0.1 | 8.5±0.4 | 12.7±8.2 | 0.14 |
| G2.54+0.20 | 3 | 1L | 3.6±0.2 | 8.4±0.5 | 15.9±2.2 | 0.21 |
| G2.54+0.20 | 3 | 2L | 0.7±0.1 | 11.1±2.0 | 12.7±4.7 | 0.05 |
| G8.68-0.37 | 1 | 1L | 30.8±0.4 | 41.5±0.2 | 18.8±0.3 | 1.54 |
| G8.68-0.37 | 1 | 2L | 16.2±0.4 | 38.4±0.2 | 12.7±0.4 | 1.20 |
| G8.68-0.37 | 1 | 4 | 3.9±0.5 | 38.9±0.2 | 3.4±0.6 | 1.05 |
| G8.68-0.37 | 1 | 5 | 3.4±0.6 | 42.3±0.4 | 4.2±1.0 | 0.75 |
| G8.68-0.37 | 2 | 1L | 39.3±0.4 | 38.5±0.2 | 37.5±0.3 | 0.99 |
| G8.68-0.37 | 2 | 2L | 14.9±0.4 | 34.5±0.3 | 27.8±0.9 | 0.50 |
| G8.68-0.37 | 3 | 1L | 3.1±0.8 | 46.6±7.2 | 60.8±22.3 | 0.05 |
| G8.68-0.37 | 3 | 2L | 2.8±0.4 | 31.1±1.0 | 12.7±4.7 | 0.21 |
| G8.68-0.37 | 4 | 1L | -46.0±1.0 | 35.6±0.3 | 28.1±0.8 | -1.54 |
| G8.68-0.37 | 4 | 2L | -49.4±1.3 | 35.9±0.5 | 43.7±1.3 | -1.06 |
| G8.68-0.37 | 4 | 4 | -20.8±1.0 | 35.7±0.1 | 6.1±0.3 | -3.19 |
| G8.68-0.37 | 4 | 5 | -15.9±0.8 | 35.6±0.1 | 5.6±0.4 | -2.63 |
| G10.44-0.02 | 1 | 1L | 55.8±1.5 | 75.6±0.4 | 31.0±1.0 | 1.69 |
| G10.44-0.02 | 1 | 2L | 23.5±1.6 | 74.6±0.8 | 27.0±2.6 | 0.82 |
| G10.44-0.02 | 1 | 4 | 1.2±0.2 | 76.7±0.2 | 2.5±0.4 | 0.42 |
| G10.44-0.02 | 2 | 1L | 30.0±1.3 | 75.6±0.6 | 29.0±1.5 | 0.97 |
| G10.44-0.02 | 2 | 2L | 10.4±1.3 | 74.5±1.2 | 22.6±4.2 | 0.43 |
| G11.94-0.15 | 1 | 1L | 43.5±0.8 | 42.1±0.2 | 24.8±0.5 | 153.79 |
| G11.94-0.15 | 1 | 2L | 14.9±0.7 | 42.2±0.5 | 15.9±0.6 | 87.88 |
| G11.94-0.15 | 2 | 1L | 39.8±0.9 | 44.0±0.3 | 24.3±0.7 | 1.53 |
| G11.94-0.15 | 2 | 2L | 15.4±0.8 | 42.6±0.5 | 15.2±0.7 | 0.95 |
| G12.68-0.18 | 1 | 1L | 74.0±1.9 | 56.5±0.4 | 32.9±1.0 | 2.11 |
| G12.68-0.18 | 1 | 2L | 15.8±1.5 | 54.3±1.0 | 15.6±1.4 | 0.95 |
| G12.68-0.18 | 2 | 1L | 35.1±0.9 | 60.7±0.4 | 30.4±0.9 | 1.08 |
| G12.68-0.18 | 2 | 2L | 7.4±0.6 | 59.1±0.6 | 12.7±6.3 | 0.55 |
| G12.68-0.18 | 3 | 1L | -29.7±0.6 | 58.0±0.3 | 27.5±0.6 | -1.01 |
| G12.68-0.18 | 3 | 2L | -4.6±0.3 | 56.4±0.9 | 12.7±0.3 | -0.34 |
| G12.68-0.18 | 4 | 1L | 12.9±1.0 | 60.4±1.5 | 35.6±2.9 | 0.34 |
| G12.68-0.18 | 4 | 2L | 21.0±1.3 | 54.2±1.8 | 56.1±3.7 | 0.35 |
| G12.68-0.18 | 4 | 4 | 6.0±0.8 | 55.2±0.3 | 5.1±1.1 | 1.08 |
| G12.68-0.18 | 4 | 5 | 5.0±0.5 | 56.3±0.1 | 2.1±0.4 | 2.03 |
| G19.47+0.17 | 1 | 1L | 93.6±2.5 | 20.9±0.4 | 26.8±0.9 | 3.29 |
| G19.47+0.17 | 1 | 2L | 44.5±2.2 | 19.7±0.5 | 19.7±1.1 | 2.13 |
| G19.47+0.17 | 1 | 4 | 14.6±0.7 | 20.9±0.2 | 9.2±0.6 | 1.49 |
| G19.47+0.17 | 1 | 5 | 7.5±0.6 | 21.4±0.3 | 7.1±0.7 | 0.98 |

**Table 5.** Observed properties of the $NH_3$(1,1) & (2,2) low spectral resolution [12.7 kms$^{-1}$] and (4,4) & (5,5) [spectral resolution 0.8 kms$^{-1}$] transitions (denoted 1L, 2L, 4 and 5 respectively in the Transition column). $\int T_B^* dV$ is the integrated brightness temperature derived from the spectra taken at the peak of the $NH_3$(1,1) emission and $V_{LSR}$ is the velocity of the line. $\Delta V$ is the line width which has been deconvolved with the instrumental response for the higher spectral resolution $NH_3$(4,4) and (5,5) transitions (see § 2). The final column gives the peak flux output from the CLASS fit to the lines (as detailed in § 2).

| Region | Core | Transition | $\int T_B^* dV$ (K.km/s) | $V_{LSR}$ (km/s) | $\Delta V$ (km/s) | Peak (K) |
|---|---|---|---|---|---|---|
| G19.47+0.17 | 2 | 1L | 14.4±1.2 | 19.6±0.9 | 22.1±1.9 | 0.61 |
| G19.47+0.17 | 2 | 2L | 5.5±1.0 | 17.9±2.7 | 14.3±6.4 | 0.36 |
| G23.44-0.18 | 1 | 1L | 73.4±1.1 | 102.1±0.2 | 28.2±0.5 | 2.44 |
| G23.44-0.18 | 1 | 2L | 29.6±0.9 | 100.9±0.3 | 16.0±0.5 | 1.74 |
| G23.44-0.18 | 1 | 4 | 4.1±0.4 | 102.4±0.6 | 10.4±1.2 | 0.37 |
| G24.79+0.08 | 1 | 1L | 146.6±2.4 | 111.6±0.2 | 29.6±0.6 | 4.65 |
| G24.79+0.08 | 1 | 2L | 79.2±2.3 | 110.5±0.3 | 22.0±0.8 | 3.39 |
| G24.79+0.08 | 1 | 4 | 23.0±0.7 | 111.0±0.1 | 7.1±0.3 | 3.02 |
| G24.79+0.08 | 1 | 5 | 19.0±0.6 | 110.8±0.1 | 7.0±0.3 | 2.54 |
| G24.79+0.08 | 2 | 1L | 21.6±0.8 | 20.6±0.4 | 23.6±0.9 | 0.86 |
| G24.79+0.08 | 2 | 2L | 10.1±0.7 | 20.0±0.8 | 15.7±1.1 | 0.60 |
| G24.85+0.09 | 1 | 1L | 41.5±1.3 | 110.8±0.4 | 25.4±0.9 | 1.53 |
| G24.85+0.09 | 1 | 2L | 15.6±0.9 | 107.6±1.1 | 14.5±3.7 | 1.01 |
| G24.85+0.09 | 2 | 1L | 8.5±0.7 | 110.5±1.1 | 25.0±2.2 | 0.32 |
| G24.85+0.09 | 2 | 2L | 4.5±1.0 | 104.3±5.3 | 48.2±11.2 | 0.09 |
| G25.83-0.18 | 1 | 1L | 145.2±6.1 | 95.2±0.6 | 29.6±1.5 | 4.61 |
| G25.83-0.18 | 1 | 2L | 83.2±6.0 | 94.1±0.7 | 23.4±2.4 | 3.33 |
| G25.83-0.18 | 1 | 4 | 18.7±1.2 | 94.3±0.3 | 8.5±0.7 | 2.06 |
| G25.83-0.18 | 1 | 5 | 9.7±0.6 | 94.0±0.2 | 6.5±0.5 | 1.37 |
| G28.28-0.36 | 1 | 1L | 19.9±0.5 | 53.5±0.4 | 27.8±0.9 | 0.67 |
| G28.28-0.36 | 1 | 2L | 5.2±0.4 | 47.6±0.8 | 12.7±5.3 | 0.39 |
| G28.28-0.36 | 2 | 1L | 21.3±0.9 | 55.7±0.7 | 36.1±1.8 | 0.56 |
| G28.28-0.36 | 2 | 2L | 10.1±1.1 | 51.0±3.2 | 58.2±7.1 | 0.16 |
| G28.28-0.36 | 3 | 1L | 32.4±0.8 | 49.2±0.4 | 31.9±0.9 | 0.95 |
| G28.28-0.36 | 3 | 2L | 3.2±0.4 | 48.9±1.6 | 12.7±1.0 | 0.24 |
| G28.28-0.36 | 4 | 1L | 26.4±0.5 | 50.7±0.3 | 33.2±0.7 | 0.75 |
| G28.28-0.36 | 4 | 2L | 5.3±0.6 | 54.8±1.9 | 34.9±4.6 | 0.14 |
| G28.28-0.36 | 5 | 1L | 12.4±0.7 | 51.6±0.7 | 25.3±1.8 | 0.46 |
| G28.28-0.36 | 5 | 2L | 3.6±0.7 | 51.3±2.0 | 20.8±4.3 | 0.16 |
| G29.87-0.04 | 1 | 1L | 18.6±1.0 | 103.1±0.6 | 20.0±1.3 | 0.87 |
| G29.87-0.04 | 1 | 2L | 5.5±0.9 | 99.3±3.1 | 13.0±14.4 | 0.40 |
| G29.96-0.02 | 1 | 1L | 130.0±2.8 | 99.2±0.2 | 29.1±0.7 | 4.20 |
| G29.96-0.02 | 1 | 2L | 100.7±3.3 | 97.3±0.5 | 36.6±1.6 | 2.58 |
| G29.96-0.02 | 1 | 4 | 21.9±0.9 | 98.6±0.2 | 7.7±0.4 | 2.67 |
| G29.96-0.02 | 1 | 5 | 21.9±0.9 | 98.4±0.2 | 8.4±0.4 | 2.43 |
| G29.98-0.04 | 1 | 1L | 84.6±1.2 | 103.0±0.2 | 25.5±0.5 | 3.12 |
| G29.98-0.04 | 1 | 2L | 24.8±1.0 | 100.5±0.5 | 15.8±0.7 | 1.48 |
| G30.79+0.20 | 1 | 1L | 89.7±2.5 | 84.5±0.4 | 32.0±1.0 | 2.63 |
| G30.79+0.20 | 1 | 2L | 28.5±1.9 | 81.1±0.6 | 16.9±2.8 | 1.59 |
| G31.28+0.06 | 1 | 1L | 47.5±2.1 | 112.1±0.5 | 23.0±1.2 | 1.94 |
| G31.28+0.06 | 1 | 2L | 27.5±1.7 | 109.6±1.0 | 14.2±6.1 | 1.82 |
| G31.28+0.06 | 1 | 4 | 5.7±0.7 | 110.4±0.3 | 5.6±0.7 | 0.95 |
| G31.28+0.06 | 1 | 5 | 5.4±0.9 | 110.5±0.6 | 7.8±1.7 | 0.64 |

**Table 5.** Continued.

respectively are lower than expected from UCHII regions ($n_e \geqslant 10^4\,\mathrm{cm}^{-3}$ and EM$\geqslant 10^7\,\mathrm{pc\,cm}^{-6}$ (Kurtz 2005)). However, care must be taken when interpreting the nature of the continuum sources from these derived values (Table 3). With a synthesised beam width $\sim$8$''$, we can resolve objects down to linear sizes $\sim$0.2 pc at a typical source distance of 5kpc – twice the upper size limit of a canonical hot molecular core. However, compared to other continuum studies, which typically have sub-arcsecond resolution, the much larger beam size will mean a large underestimate of electron density and emission measure (Eq. 2, 3 and 4). For example, if the source is 0.02 pc instead of 0.2 pc then the electron density and emission measure increase by factors of around 30 and 100 for a given flux density.

With this in mind, a possible explanation for the detection of sources at 24 GHz but not 8 GHz may be that the continuum emission is optically thick rather than optically thin between 8 and 24 GHz, despite their seemingly low electron densities. In the optically thick regime, the spectral index ($S_\nu \propto \nu^\alpha$) changes from $\alpha = -0.1$ to $\alpha = 2$. If the source flux at 8 GHz lies just below the detection limit of $\sim$5mJy (Walsh et al. 1998), the 24 GHz fluxes of $\geqslant$50mJy for some sources would be consistent with this. An alternative possibility is that the 24 GHz continuum sources were missed by Walsh et al. (1998) as they are too extended and were spatially filtered or resolved-out by the more extended array configuration they used. However, it is not currently possible to differentiate between the two scenarios to determine the nature of the sources. Higher resolution, higher frequency observations and modelling of the different $uv$-coverage between all the observations is required to resolve this unambiguously.

### 4.3 Association of $NH_3$(1,1), methanol maser and continuum emission

The cores were split into four subgroups based on the observed combinations of $NH_3$(1,1), 6.7 GHz methanol maser and 24 GHz continuum emission as follows: isolated $NH_3$ cores (Group 1); $NH_3$ cores associated with only methanol maser emission (Group 2); $NH_3$ cores associated with 24 GHz continuum and methanol maser emission (Group 3) and $NH_3$ cores associated with 24 GHz continuum emission only (Group 4). The cores were distributed with 16, 16, 6 and 2 cores in Groups 1 to 4 respectively. Based on these four groups, most of the $NH_3$(1,1) cores are coincident with methanol maser emission (Groups 2 & 3), but there are a substantial fraction of $NH_3$ cores with neither 24 GHz continuum nor maser emission (Group 1). The properties of cores within each of the groups are discussed in §5.2.

#### *4.3.1 Selection Effects/Observational Bias*

Having found the cores can be observationally seperated into groups based on their association with $NH_3$, methanol maser and continuum emission, we are interested in investigating whether there are any inherent differences in the physical properties between the cores in the groups. However, we must be careful to ensure that this distribution of cores is real, rather than an effect of observational bias or selection effects.

It may be plausible, for instance, that undetected $NH_3$ cores, continuum emission or methanol masers exist toward the regions but lie beneath the sensitivity limits. Alternatively, we might be missing maser emission if the beam is not aligned along our line of sight. It is therefore possible that cores with fewer tracers (e.g. $NH_3$ or continuum only) may, on further inspection, move into a group with more tracers (e.g. $NH_3$ + methanol + continuum). It should be remembered that any conclusions drawn about cores within the groups are limited by the observational parameters used to define the groups. However, each of the methanol maser, $NH_3$ and continuum observations have the same sensitivity limit towards all the regions so the relative flux densitites of the tracers in each region are directly comparable.

Another possible bias might be induced by the factor of $\sim$5 variation in distance towards the regions. The poorer spatial resolution towards the most distant regions could make it more difficult to resolve emission into individual cores. If this were the case we would expect to see it manifested in two ways: firstly, as an increased number of cores detected toward the closest regions; and secondly, groups with fewer tracers (e.g. $NH_3$ or continuum only) should be more prevalent toward nearby regions than cores with multiple tracers (e.g. $NH_3$ + methanol + continuum) which cannot be resolved as seperate cores for the most distant regions. Figure 3[a] shows the number of cores per region as a function of distance to each region (G28.28-0.36 has been removed due to ambiguity over the number of reliable cores in the region as explained in § 4.4). There is no clear correlation between the detected number of cores and the distance to the region. Similarly, Figure 3[b] gives the distance to each core vs their association with $NH_3$, methanol maser and cm-continuum emission as outlined in § 4.3. The distribution of core distances between the groups appears similar and evenly spread across the whole distance range. The mean distances are 4.5±1.3, 5.1±1.6, 5.5±1.7 and 3.4±2.1 kpc for Groups 1 to 4 respectively. Although there appears to be a slight increase from Group 1 to 3, the difference is too small to have a significant affect on the resolution. With only two cores, the results from Group 4 are unreliable. Again the distance to a region appears to have little effect on the association of the star formation tracers. From this we conclude that the factor of $\sim$5 range in the resolution is not likely to bias the distribution of cores in the groups.

Finally, although the criteria outlined in §4 unambiguously separate the majority of cores into their respective groups, it should be noted that one ammonia core (G29.87-0.04 core 1) is marginally offset from the methanol maser emission and three cores not classified as associated with 24 GHz continuum (G19.47+0.17 core 1, G24.85+0.09 core 2 and G31.28+0.06 core 1) lie close to separation distance of the selection criteria for them to be associated with the continuum emission. With different association criteria these cores may have been placed in different groups.

### 4.4 The Regions

The following sub-sections give a brief summary of the individual ammonia/continuum cores detected toward each region and their association with methanol maser emission. We defer detailed analysis of individual regions for a subsequent paper. Regions with poor image fidelity or potential artifacts due to the snapshot imaging mode used for the observations are highlighted.

#### *4.4.1 G316.81-0.06*

There are two sources of ammonia emission spatially separated by $\sim$20″. The brighter of the two in the $NH_3$(1,1) transition (core 1) is seen only in emission and is not associated with either methanol maser or continuum emission. Core 2 is associated with both methanol maser and continuum emission and has both an emission and weak absorption component in the $NH_3$(1,1) transition (cores 2a and 2b respectively). The absorption is offset by 3 $kms^{-1}$ from the emission core suggesting the continuum source at the maser position is not associated with core 2a. Core 2b is only seen in the $NH_3$(1,1) transition so is not included in further analysis. There is an unresolved (4,4) and weak (5,5) detection at the maser site and velocity of the emission core. The $NH_3$(4,4) emission shows some asymmetry suggesting the absorption from core 2b is present there too.

The continuum source at the maser position was not detected in previous surveys and the nature of the emission is discussed in §4.2. The second continuum source, offset $\sim$ 30″ South West, has a cometary morphology and was previously detected at 8GHz. The extended morphology and position towards the edge of the primary beam make it difficult to derive accurate flux and image sizes using fitting procedures in both image and $uv$ planes. Fitting multiple Gaussian profiles did not improve the fitting.

#### *4.4.2 G323.74-0.26*

There is a single ammonia core extended north-south with strong $NH_3$(1,1) hyperfine asymmetries and a strong, unresolved $NH_3$(4,4) and (5,5) detection toward the methanol maser position.

#### *4.4.3 G331.28-0.19*

There is a single ammonia core associated with the methanol maser position with strong $NH_3$(4,4) and (5,5) emission. There is a weak, extended continuum source coincident with the methanol maser not previously detected at 8GHz. The nature of sources with continuum detected at 24GHz but not 8GHz is discussed in §4.2.

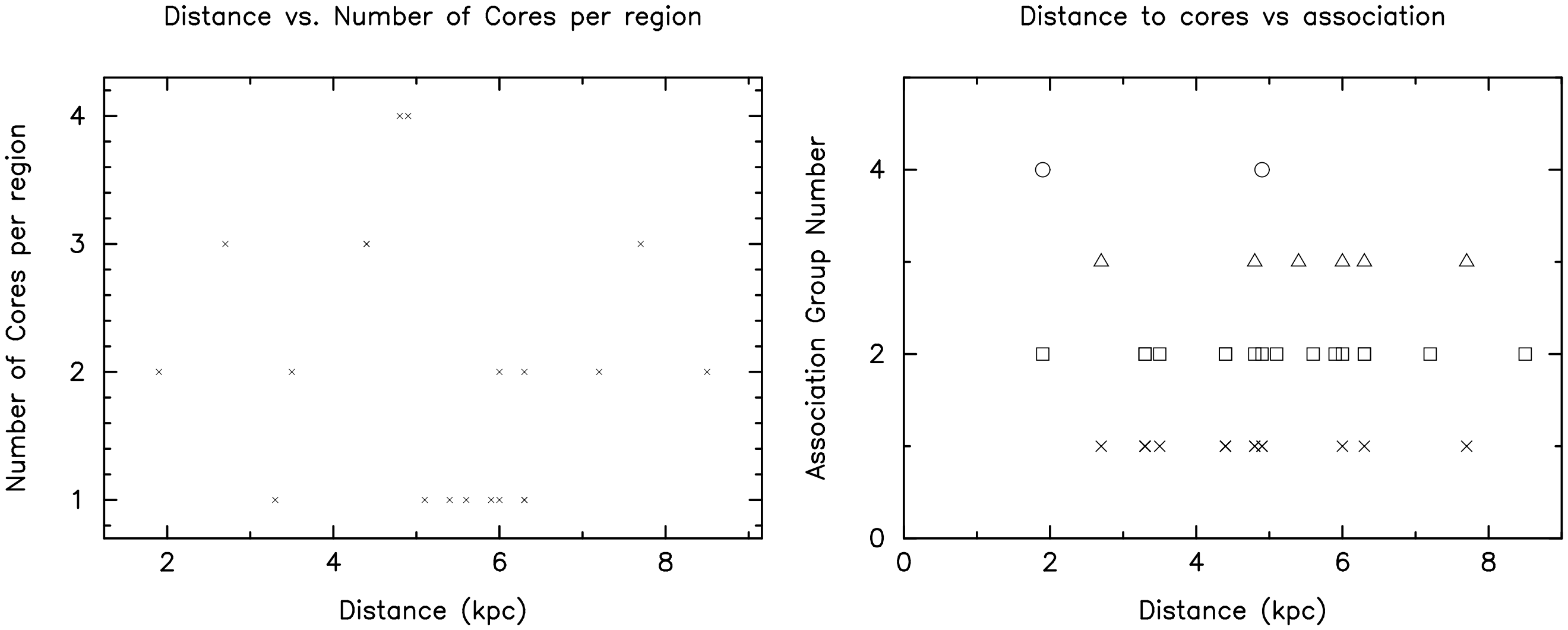

**Figure 3.** [a](Left) Distance to each region vs the total number of cores (including continuum) detected towards the region. G28.28-0.36 has been removed due to ambiguity over the number of reliable core detections. There appears to be no correlation between the number of cores detected and the distance to the region. [b] (Right) Distance to each core vs their association with $NH_3$, methanol maser and cm-continuum emission as outlined in § 4.3: isolated $NH_3$ cores (Group 1, crosses); $NH_3$ cores associated with only methanol maser emission (Group 2, squares); $NH_3$ cores associated with 24GHz continuum and methanol maser emission (Group 3, triangles) and $NH_3$ cores associated with 24GHz continuum emission only (Group 4, circles). Again the distance to a region appears to have little effect on the association of the star formation tracers.

#### *4.4.4 G332.73-0.62*

There are two extended ammonia cores in the field separated by $\sim40''$, the brighter of which in $NH_3(1,1)$ peaks at the methanol maser position. The morphology of the cores is filamentary from North-East to South-West.

#### *4.4.5 G1.15-0.12*

At the methanol maser position is a single slightly resolved ammonia core. The $V_{LSR}$ of the maser emission lies slightly outside the measured FHWM of the ammonia $V_{LSR}$ so strictly they should not be associated, given the criteria outlined in §4. However, the difference is only equivalent to a single channel ($0.2 kms^{-1}$) so the ammonia emission has been considered associated with the methanol maser emission for the following analysis. There is a continuum source offset 90" south west, outside the primary beam. This has previously been detected at 8GHz with no coincident methanol maser emission. It is difficult to tell whether there is any ammonia emission at the continuum site due to sidelobe structure and noise amplification from primary beam correction.

#### *4.4.6 G2.54+0.20*

Extending North-East to South-West is a single ammonia filament with three individual cores, each separated by $\sim20''$. The central core is associated with the methanol maser and has the largest line widths of $2.6\,kms^{-1}$.

#### *4.4.7 G8.68-0.37*

Complicated structure and a strong continuum source make the field difficult to image. As a result, possibly exacerbated by missing flux, there is a strong negative sidelobe in between the two maser positions. There are four cores in total, with a large extended ammonia clump seen in emission surrounding the first maser site (core 1) with unresolved $NH_3(4,4)$ and (5,5) emission at the site of the methanol maser. The second maser site, offset $\sim1'$ South-West, is coincident with a strong continuum source and ammonia core in absorption from $NH_3(1,1)$ to (5,5) (core 4). Ammonia is seen in emission surrounding core 4, with an extended, very optically thick core to the North and a weaker component to the South (cores 2 and 3). The low resolution $NH_3(1,1)$ and (2,2) spectra for core 3 have very different line profiles: the $NH_3(1,1)$ emission appears much broader than the sharply peaked $NH_3(2,2)$ profile. However, this may simply be due to the low signal to noise of the spectrum.

#### *4.4.8 G10.44-0.02*

A single ammonia filament extends North-South through the field with two core peaks, the brightest of which in $NH_3(1,1)$ is coincident with the methanol maser emission and weak, unresolved $NH_3(4,4)$ emission. The second peak, $20''$ North, is offset by $\sim2.2\,kms^{-1}$ in velocity with a slightly smaller line width (2.0 vs $2.6\,kms^{-1}$). No continuum is seen towards the region.

#### *4.4.9 G11.94-0.15*

In the field are two ammonia cores separated by $30''$ spatially and $\sim2\,kms^{-1}$ in velocity. The brightest of the two in the $NH_3(1,1)$ transition (core 1) is co-incident with the methanol maser. The outer hyperfine component of core 2 shows a minor, very narrow peak on top of the main, broader peak. As this is not replicated in the other hyperfine components, it is likely to be an artifact.

There is a second methanol maser $\sim2'$ south-west, outside the primary beam, co-incident with an unresolved continuum source previously detected at 8GHz (core 3). It is difficult to tell whether there is any ammonia emission at the continuum site due to sidelobe structure and noise amplification from primary beam correction.

#### *4.4.10 G12.68-0.18*

Four ammonia cores are detected in the region extending North-East to South-West. The outer two cores (1 and 2), seen in emission, are brightest in the $NH_3$(1,1) transition and very optically thick. Core 2 shows strong $NH_3$(1,1) line profile asymmetries and much narrower line width than the more extended core 1. Core 3 is close to the maser emission and seen in absorption at the site of a strong continuum source, not previously detected at 8GHz. The nature of the continuum source is discussed in §4.2. Core 4 is also close to the maser emission and has unresolved $NH_3$(4,4) and (5,5) emission. Comparison of the core velocities shows that core 4 is associated with the methanol maser emission. The nature of this core is discussed in more detail in §4.1.

#### *4.4.11 G19.47+0.17*

There are two ammonia cores in the region separated by slightly more than a synthesised beam width. The brightest, at the methanol maser position, has very broad line widths with possibly two slightly resolved velocity components. Weak, unresolved $NH_3$(4,4) and (5,5) emission is also seen coincident with the methanol maser emission. The second ammonia peak, with a much smaller line width, is coincident with a continuum source $\sim 15''$ north-west of the methanol maser position. No continuum was previously detected at 8 GHz at this position.

#### *4.4.12 G23.44-0.18*

The single, ammonia core is extended North-South, in the same orientation as the two methanol masers. There is a velocity gradient along the core of $\sim 2\,\mathrm{kms^{-1}}$ with the line width increasing towards the southern maser position. The integrated intensity $NH_3$(4,4) image shows emission at both methanol maser positions but only a weak spectra can be extracted at the southern maser site. No continuum emission is seen towards the region.

#### *4.4.13 G24.79+0.08*

There are two continuum sources in the field, one at the methanol maser position (core 1) the other offset $1'$ North-West at the edge of the primary beam (core 3). Neither of the sources were seen at 8GHz. The ammonia in emission at core 1 has a large line width with strong, unresolved $NH_3$(4,4) and (5,5) emission at the methanol maser position. The second ammonia core, $\sim$10$''$ North, has a much narrower line width. No ammonia is detected at core 3 although this may be due to noise amplification from primary beam correction.

#### *4.4.14 G24.85+0.09*

Both continuum and ammonia emission are seen to peak near the methanol maser position in this region (core 1). They are extended in roughly the same direction, South-East, but may be slightly offset from each other. The ammonia $NH_3$(1,1) emission shows large profile asymmetries. The hyperfine transitions of the second ammonia core, offset by 20$''$, are only just discernible above the noise.

#### *4.4.15 G25.83-0.18*

There is a single ammonia core with strong emission from $NH_3$(1,1) to (5,5) and large linewidth associated with the methanol maser. The unresolved $NH_3$(4,4) and (5,5) emission peaks at the methanol maser position. Hyperfine components are seen in the $NH_3$(4,4) spectra showing this transition must be quite optically thick.

#### *4.4.16 G28.28-0.36*

Due to complicated source structure, a strong continuum source and a poor synthesised beam response (as the source is nearly equatorial) the region is very difficult to image. There appears to be strong ammonia emission outside the primary beam but its position is uncertain. The best CLEAN image was produced with sources at the position of cores 3, 4 and 5. The characteristic spectra taken from these cores are very similar so it is unclear if they really are independent cores or sidelobes of another source outside the primary beam. As a result, they have not been included in further analysis.

There are two ammonia cores in emission, within the primary beam, close to the maser position (core 1 & 2). It is unclear which is associated with the maser within the pointing uncertainty. Core 1 has a very narrow line width and core 2 is very optically thick with extreme $NH_3$(1,1) hyperfine asymmetries. However, the $V_{LSR}$ of both these cores lies outside their measured FWHM of the maser's $V_{LSR}$. As the velocity difference is small ($<$1kms$^{-1}$) the methanol maser has still been associated with core 1 which is closer both spatially and in velocity than core 2. Interestingly, ammonia is seen strongly in absorption at the position of the maser emission with the correct velocity. However, this is at the same velocity as core 5, which is seen in emission, so with relatively poor image quality may in fact be a negative sidelobe of emission from this core. Therefore the absorption component has not been included as a separate core in the analysis. Higher fidelity imaging may resolve this issue. The isolated continuum source (core 6) 30$''$ North-East of the maser emission was previously detected at 8GHz.

#### *4.4.17 G29.87-0.04*

There is a single, bright, slightly extended ammonia core in emission which peaks $\sim$10$''$ South of the methanol maser position. Although this distance equals the synthesised beam size (the limit of the selection criteria), the ammonia and methanol emission have been considered associated as the low lying ammonia emission extends towards the methanol maser. The channel map shows more complicated source structure at lower emission levels but it it unclear if this is real or due to sidelobes of the main core. The high-resolution $NH_3$(1,1) spectra shows a slight absorption dip blue-ward of the main hyperfine component. However, the fact that the spectra is noisy and no other absorption dips are seen blue-ward of the other hyperfine components suggests this is an artifact.

#### *4.4.18 G29.96-0.02*

A single continuum and extended ammonia core are seen in emission near the methanol maser position. The ammonia $NH_3$(1,1) emission contains two components (1a and 1b) separated in velocity by $\sim 2\,\mathrm{kms^{-1}}$. The first component is seen over the full extent of the $NH_3$(1,1) emission while the second is only seen at the methanol maser position at the same position and velocity as very strong $NH_3$(4,4) and (5,5) emission. There is no corresponding peak in the $NH_3$(4,4) and (5,5) spectra at the 1a velocity. We interpret this as either internal heating (i.e. a hot, unresolved component at the methanol maser position [1b] surrounded by a cooler envelope [1a]) or a chance line of sight effect.

#### *4.4.19 G29.98-0.04*

There is a single ammonia core extended North-South associated with the methanol maser emission. A slight velocity gradient is seen from East to West across the core. The line width and $NH_3$(1,1) line profile asymmetry increases toward the centre from all directions. No continuum emission is seen towards the region.

#### *4.4.20 G30.79+0.20*

There is a single extended ammonia core which peaks $\sim$10$''$ South of the methanol maser position. Although this distance equals the synthesised beam size (the limit of the selection criteria), the ammonia and methanol emission have been considered associated as the low lying ammonia emission extends towards the methanol maser. The ammonia line width increases toward the central peak. No continuum emission is seen towards the region.

#### *4.4.21 G31.28+0.06*

Due to complicated source structure, a strong continuum source and a poor synthesised beam response (as the source is nearly equatorial) the region is difficult to image. There appears to be ammonia emission outside the primary beam but its position in uncertain. Within the primary beam there is an ammonia core extended East-West with a velocity gradient of $\sim$3 kms$^{-1}$ along the filament. The peak of the $NH_3$(1,1) emission shows unresolved weak $NH_3$(4,4) and possibly (5,5) detections. An unresolved continuum source, previously detected at 8GHz, lies $\sim$10$''$ South-West of the maser emission. Using the conditions outlined in §4, it is not possible to unambiguously separate the ammonia emission into cores and assign their association with methanol maser and continuum emission. Given the continuous velocity gradient, the whole filament has been assigned as a single core associated with the methanol masers (core 1) which is offset from the continuum source (core 2).

## 5 ANALYSIS

### 5.1 Properties of $NH_3$ cores

We now show the distribution of the observed core properties in comparison with those from other molecular line observations towards high mass star forming regions, which are shown in Table 6. There is little overlap between the observed sources in this work and those in the studies listed in Table 6.

#### *5.1.1 $NH_3$(1,1) Core Size*

Figure 4(a) plots a histogram of the measured core sizes from the $NH_3$(1,1) emission. The core sizes range from 0.09 to 0.58 pc with a mean value and standard deviation of 0.28$\pm$0.13 pc. The drop in the number of cores at diameters $<$0.2 pc reflects the completeness limit due to the angular resolution of the observations. Similarly, the cutoff at the high diameter end may be due to spatial filtering of structures larger than that sampled by the shortest baseline (see § 4). However the slope between 0.2 and 0.6 pc is completely sampled and shows a steep drop off at larger core diameters.

The measured sizes are significantly smaller than the ammonia cores reported by Wu et al. (2006) of 0.4 to 3.1 pc and the mean size of 0.57 pc reported by Pillai et al. (2006) toward infrared dark clouds. This is likely due to the larger beam size used ($\sim$40$''$) and the fact their single dish observations do not suffer from spatial filtering. The cores are in fact closer in size to those of the CS observations of Cesaroni et al. (1999) and Fontani et al. (2005).

#### *5.1.2 $NH_3$(1,1) Linewidth*

Figure 4(b) shows the histogram of the $NH_3$(1,1) linewidths derived from the fits to the characteristic spectra. As discussed in §4.1, cores with potentially blended profiles have been removed from the analysis. The remaining linewidths vary from 0.7 to 4.6 kms$^{-1}$ with a mean and standard deviation of 1.85$\pm$0.95 kms$^{-1}$. The minimum measured linewidth is well above the spectral resolution (0.2 kms$^{-1}$) so the cutoff is a true reflection of the distribution. If the cores obey the linewidth-size relation (e.g. Larson 1981, which is discussed further in §5.1.4), the cutoff may instead be caused by the finite spatial resolution of the observations. The distribution is weighted towards cores with smaller linewidths, a trend which is reinforced if the different velocity components of the blended spectra are treated as individual cores.

The mean line width of 1.85 kms$^{-1}$ is lower than those of 3.9 and 4.5 kms$^{-1}$ for $HCO^+$ and $CH_3CN$ respectively, found by Purcell et al. (2006) toward the same regions. This may be due to the smaller beam size resolving structure integrated over the $\sim$35$''$ Mopra beam used by Purcell et al. (2006). Alternatively the different molecules may be tracing different gas [$HCO^+$ is known to be enhanced in outflows for example (e.g. Rawlings et al. 2004)]. The line widths are comparable to the $NH_3$ cores reported by Wu et al. (2006), Pillai et al. (2006) and Sridharan et al. (2002) despite their larger beam size of $\sim$40$''$. These linewidths are significantly smaller than those in Table 6 towards UCHII regions, most likely due to increased turbulent injection from outflows etc. in the more evolved sources.

#### *5.1.3 $NH_3$(1,1) Optical Depth*

Figure 4(c) shows a histogram of the $NH_3$(1,1) optical depth, $\tau_{(1,1)}$, derived from the characteristic spectra, which varies from 0.3 to 18.3 with a mean and standard deviation of 3.9$\pm$3.4. With the exception of 5 outliers, all the cores have optical depths less than $\sim$5. As can be seen in Figure 14, the outliers all have extreme asymmetries in their hyperfine structure which makes the derived $\tau_{(1,1)}$ values unreliable. Interpreting the asymmetries in the hyperfine profiles is discussed further in §5.4.

#### *5.1.4 Relationships between $NH_3$(1,1) Linewidth, Optical Depth and Core Size*

Figure 5 plots the measured core properties ($NH_3$(1,1) linewidth, core size and optical depth) against each other after removing possibly blended lines (see §4.1) and cores with erroneous optical depths due to large hyperfine asymmetries (see §5.1.3). The cores are separated into the groups outlined in §4 but discussion of the properties within these groups is continued in §5.2. The average errors in the fit to the linewidth and optical depth are ∼0.06 kms$^{-1}$ and 0.27 respectively, while the error in the core size is estimated to be ∼10% due to the distance uncertainty.

| Paper | Lines | Beam size ($''$) | Mean $\Delta V$ (kms$^{-1}$) | Mean core size (pc) | Notes |
|---|---|---|---|---|---|
| Churchwell et al. (1990) | $NH_3$ | 40 | 3.1 | ‡ | 70% detection rate toward UCHII regions |
| Cesaroni et al. (1991) | $C^{34}S$ | >12 | 6 | 0.3-0.5 | UCHII regions |
| Bronfman et al. (1996) | CS | 50 | 1.0-15.8 | ‡ | |
| Molinari et al. (1996) | $NH_3$ | 40 | 1.81/1.72* | ‡ | |
| Cesaroni et al. (1999) | CS/$C^{34}S$ | 29/13 | 3.4/3.1 | 0.22/0.21 | Pre-UCHII regions ("High" *) |
| Hofner et al. (2000) | $C^{17}O/C^{18}O$ | 11-22 | 6.22/5.85 | 1 | UCHII regions |
| Brand et al. (2001) | CS/$C^{34}S$† | 28/13 | 2.9/2.0 | 0.75/0.44 | Pre-UCHII regions ("Low" *) |
| Molinari et al. (2002) | $HCO^+$† | 4-9 | 2.6 | 0.3-1 | |
| Sridharan et al. (2002) | CO/$NH_3$ | 11/40 | 2.0 | ‡ | |
| Beuther et al. (2002) | CS | 11-27 | 3.0 | | |
| Fontani et al. (2005) | CS/$C^{17}O$ | 20/50 | 2.7/2 | 0.1-2 | $\delta < -30°$ ("Low" *) |
| Wu et al. (2006) | $NH_3$ | 40 | 1.54 | 1.6 | 46% detection rate toward $H_2O$ masers |
| Pillai et al. (2006) | $NH_3$ | 40 | 1.7 | 0.57 | |
| Purcell et al. (2006) | $HCO^+/CH_3CN$† | 35 | 3.9/4.5 | ‡ | |
| This work | $NH_3$ | 8/11 | 1.85 | 0.28 | |

**Table 6.** Properties of high mass cores from other observational surveys. Only the papers with molecular line observations are included. Where possible, values have been included for each of the molecular species and different groups within the survey (see below). † indicates that other molecular lines were also observed but have not been included in the table. ‡ indicates single pointing observations so no core size is available. * The sources for several surveys are split into two groups, named "High" and "Low", based on their IRAS colours. The observations suggest that the "High" sources are more evolved than the "Low" sources.

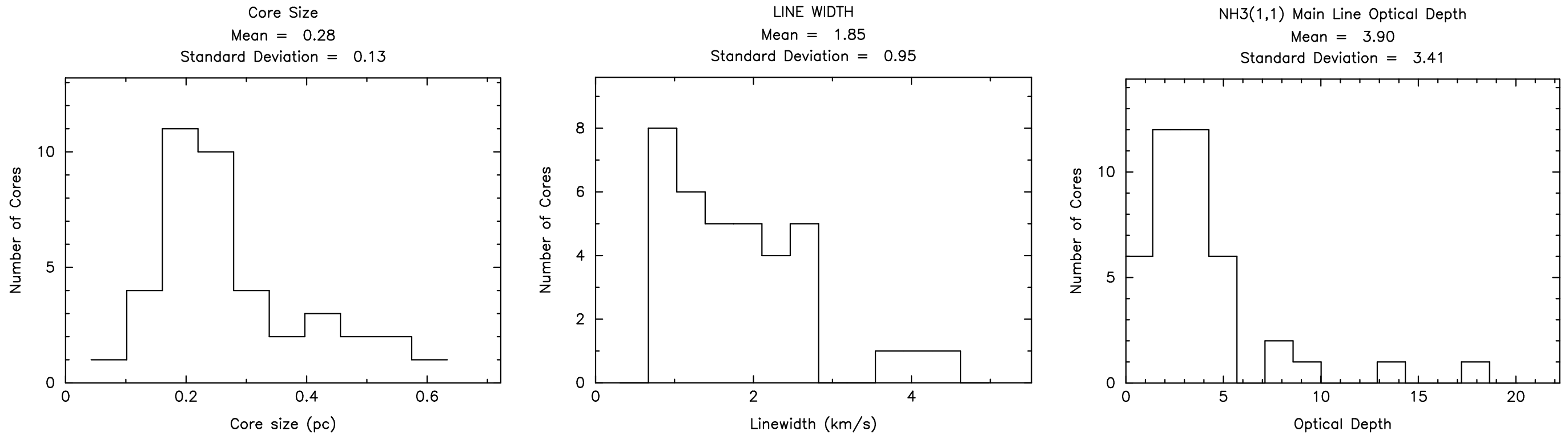

**Figure 4.** Histograms of: (a) beam-deconvolved $NH_3$(1,1) core sizes (left); (b) velocity deconvolved $NH_3$(1,1) linewidths (centre); (c) $NH_3$(1,1) main line optical depths (right).

Figure 5 shows a weak correlation between the $NH_3$(1,1) linewidth and core size. The dashed line shows the Larson (1981) linewidth-size relation for molecular clouds ($\Delta V$ = 1.1 R$^{0.38}$, with units: [$\Delta V$] = kms$^{-1}$and [R] = pc). All the observed cores have a greater velocity dispersion for a given core size than the Larson (1981) relation. The fact the correlation is weak may not be surprising if we are observing cores at a variety of evolutionary stages: more evolved sources at a given core size will have increased linewidths due to turbulence injection. There is a better correlation between the $NH_3$(1,1) linewidth and optical depth: cores with smaller linewidths have a larger optical depth. No correlation is seen between the $NH_3$(1,1) optical depth and core size.

### 5.2 $NH_3$(1,1) properties vs association with 24 GHz continuum and methanol maser emission

Having separated the $NH_3$ cores into four groups based on their association with 24 GHz continuum and methanol maser emission, we now investigate how the core properties vary with these associations. Table 7 shows the average core properties for each of the four groups. The measured linewidths vary significantly between the Groups, increasing from Groups 1 to 3 before falling slightly in Group 4. The same trend is seen after removing the potentially blended lines. The linewidth vs size plot in Figure 5 illustrates this further; although the numbers within each group are too small to perform a statistical analysis, the Group 3 sources (triangles) have noticeably larger linewidths, and Group 1 sources sit to the lower-left, for a given core size than the other groups. The Group 1 optical depths stand out in Table 7 as being twice as large as those in Groups 2 to 4. However, these values also include the cores with anomalous optical depths due to large hyperfine asymmetries. Interestingly, after removing these sources with anomalous optical depths, only the average value for Group 1 is altered (to 3.06±1.29): i.e. all the sources with the strongest profile asymmetries belong to Group 1. Finally, there is little difference in the core sizes between the groups given the small number of sources in Group 4.

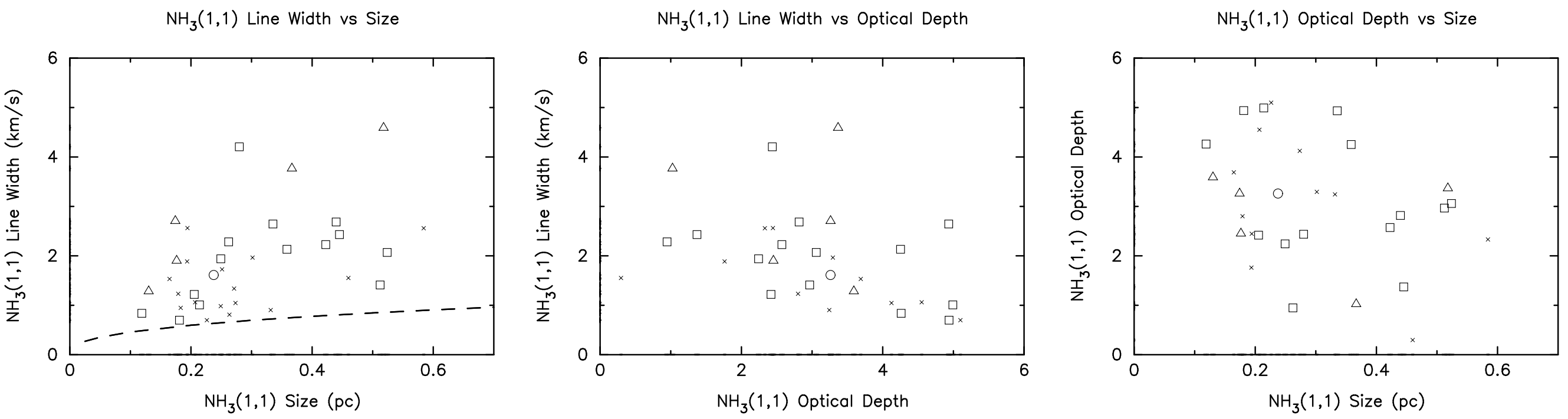

**Figure 5.** Correlation between $NH_3(1,1)$ linewidth, core size and main line optical depth: (Left) Linewidth vs Core Size, (Centre) Linewidth vs Optical Depth and (Right) Optical Depth vs Core Size. Both cores with potentially blended lines (see §4) and those with anomalously large optical depths due to hyperfine asymmetries (see §5.1.3) have been removed. The cores are separated into the groups outlined in §4 with Group 1 ($NH_3$ only), Group 2 ($NH_3$ + methanol maser), Group 3 ($NH_3$ + methanol maser + 24 GHz continuum) and Group 4 ($NH_3$ + 24 GHz continuum) shown as crosses, squares, triangles and circles respectively. The average errors in the fit to the linewidth and optical depth are 0.06 $\mathrm{kms}^{-1}$ and 0.27 respectively. The error in the core size is estimated to be 10% due to the uncertainty in the distance. The dashed line on the linewidth ($\Delta V$) vs core size (R) plot (Left) shows the Larson (1981) linewidth-size relation for molecular clouds ($\Delta V$ = 1.1 $\mathrm{R}^{0.38}$, with units: $[\Delta V]$ = $\mathrm{kms}^{-1}$ and [R] = pc). All the observed cores have a greater velocity dispersion for a given core size than the Larson (1981) relation.

| Group | Association | $\Delta V_{(1,1)}$ ($\mathrm{kms}^{-1}$) | $\tau_{(1,1)}$ | Size (pc) |
|---|---|---|---|---|
| 1 | $NH_3$ only | 1.43±0.57 | 5.69±4.62 | 0.27±0.11 |
| 2 | $NH_3$ + meth. | 2.40±1.51 | 2.94±1.33 | 0.30±0.13 |
| 3 | $NH_3$ + meth. + 24 GHz cont. | 3.00±1.15 | 2.66±0.88 | 0.27±0.13 |
| 4 | $NH_3$ + 24 GHz cont. | 2.40±0.79 | 2.78±0.48 | 0.16±0.07 |

**Table 7.** Mean and standard deviation for properties of the $NH_3$ cores in the four groups outlined in §4.

### 5.2.1 *Methanol Maser emission*

In each region, all of the methanol masers are associated both spatially and in velocity (with the possible exception G28.28-0.36) with an ammonia core, strongly suggesting they are from the same gas in the region (within the typical spatial resolution of the observations of ∼0.2 pc).

### 5.2.2 *Continuum Sources*

With the exception of G24.79+0.08 core 3 (which lies at the edge of the primary beam making an ammonia detection difficult), all of the continuum cores seen at 24 and *not* at 8 GHz, are also associated with ammonia emission. This is in contrast to the cores detected at both 8 and 24 GHz, only two of which are associated with ammonia. As molecular gas is expected to be detected toward cores still undergoing star formation, this suggests that the cores seen only at 24 GHz are younger than those seen at both 8 + 24 GHz.

## 5.3 Properties of Cores with $NH_3(4,4)$ and (5,5) emission

$NH_3(4,4)$ emission is detected toward the peak of 13 $NH_3(1,1)$ cores and 11 of these also have coincident $NH_3(5,5)$ emission. We now investigate whether these cores are inherently different from the remaining core population. Considering that the relative energy levels are populated according to a Boltzmann distribution, cores with detected $NH_3(4,4)$ emission must either be warmer or have stronger $NH_3$ emission. Figure 6 plots the integrated $NH_3(1,1)$ intensity (giving the relative strength of the $NH_3$ emission) vs the ratio of the $NH_3(2,2)$ and (1,1) integrated intensities for cores with detected $NH_3(4,4)$ as triangles and those without as crosses. There is no significant difference in the integrated $NH_3(1,1)$ intensity between cores with and without detected $NH_3(4,4)$ emission. This shows that the $NH_3(4,4)$ emission is not simply detected because the $NH_3$ emission is brighter for those cores. However, there is a clear separation in the ratio of the integrated intensities between the populations: cores with detected $NH_3(4,4)$ emission have a higher $NH_3(2,2)/(1,1)$ integrated intensity ratio. This shows that these cores are warmer than those without detected $NH_3(4,4)$ emission. A more conclusive result requires the derived temperature and core column density (a distance independent measure of the amount of $NH_3$) and will be addressed in Paper II.

### 5.3.1 *Association with 24 GHz continuum and methanol maser emission*

The higher spatial resolution of the $NH_3(4,4)$ and (5,5) compared to the $NH_3(1,1)$ observations (8″vs 11″) provides a stronger constraint to the criteria outlined in §4 as to whether this emission is associated with either methanol or continuum emission. In every case, the $NH_3(4,4)$ and (5,5) emission is unresolved and both within a synthesised beam width of the methanol maser emission spatially and within the FWHM in velocity. As discussed in §4.4, the $NH_3(1,1)$ emission of the outlier in Group 4, G31.28+0.06 core 1, is only marginally outside the criteria of being associated with the methanol maser emission. However, the $NH_3(4,4)$ emission is clearly centred on the methanol maser.

In comparison to the four groups outlined in §5.3.1, the cores with detected $NH_3$(4,4) emission are split 8, 5 into Groups 2 and 3 respectively. In other words, the $NH_3$(4,4) emission is only associated with cores which also have maser/continuum emission – there is no $NH_3$(4,4) detected towards cores with only $NH_3$(1,1) emission. Combined with the result from Figure 6, this points to the fact that the cores with maser/continuum emission are also warmer than those with only $NH_3$ emission.

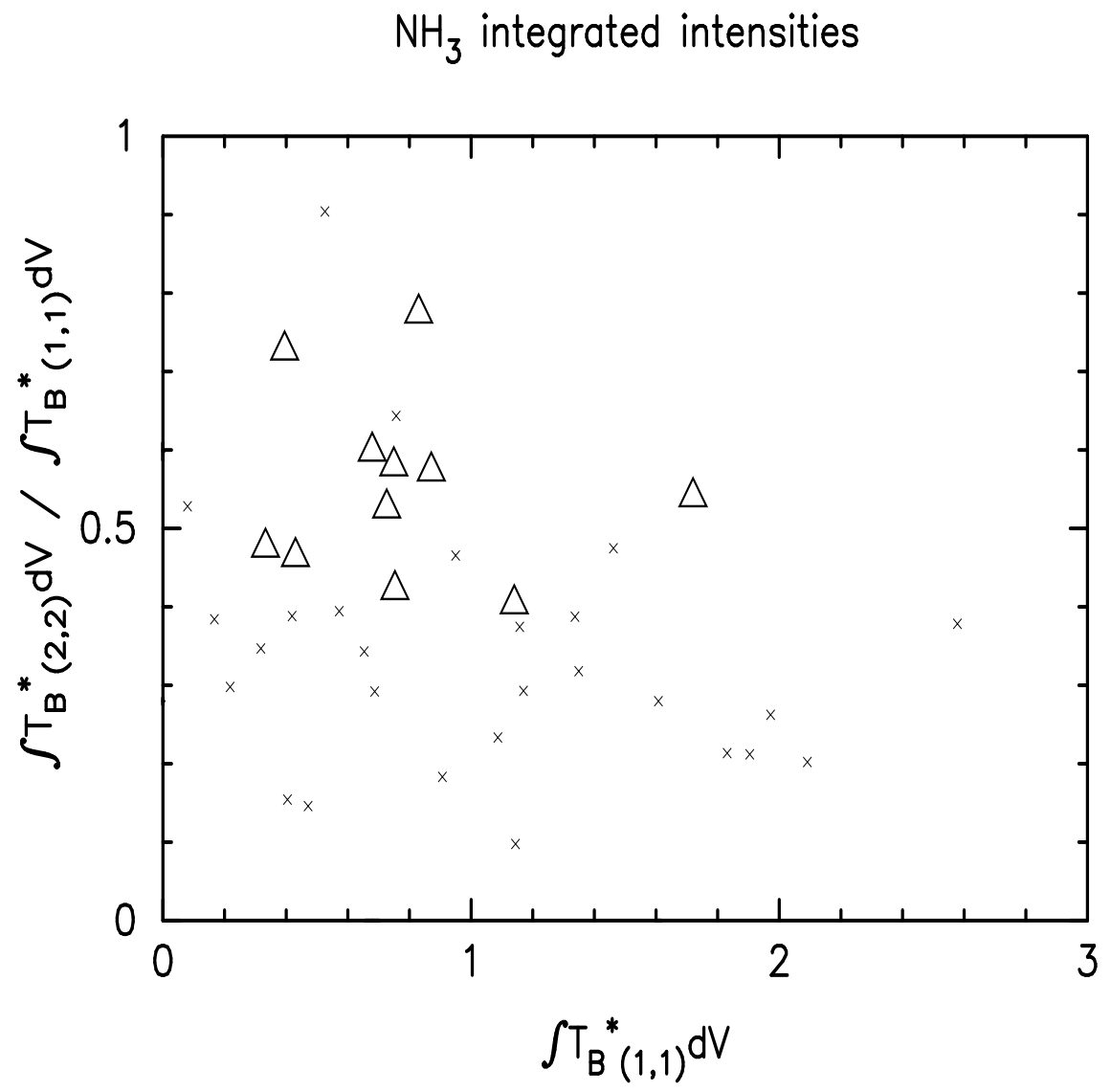

**Figure 6.** Integrated $NH_3$(1,1) intensity vs the ratio of the $NH_3$(2,2) and (1,1) integrated intensities. Cores with detected $NH_3$(4,4) are shown as triangles and those without are shown as crosses. While there is no significant difference in the integrated $NH_3$(1,1) intensity between the populations, the cores with detected $NH_3$(4,4) emission clearly have a higher $NH_3$(2,2) to (1,1) ratio and are therefore warmer than cores without detected $NH_3$(4,4) emission.

### 5.3.2 *Linewidth vs Transition Energy*

Measurements of the linewidth from several transitions provides additional information on the kinematic state of the gas in the cores. Figure 7(a) plots the $NH_3$(4,4) linewidths (from Table 5) and high spectral resolution $NH_3$(1,1) linewidths (from Table 4) for all cores with detected $NH_3$(4,4) emission. Although the observations were taken with different spectral resolution [shown as dashed lines in Figure 7(a)] all the lines are well resolved. It is clear that most of the $NH_3$(4,4) linewidths are significantly broader than those of the $NH_3$(1,1) emission. Although these cores are found to be warmer, the increase in linewidth cannot be solely due to the temperature – even at an extreme temperature of 2000K, at which the molecular gas and dust should have been destroyed, the thermal broadening is only $\sim$2.3 kms$^{-1}$. Instead, there must be additional broadening in the warmer gas with respect to the colder gas, either due to systematic motion (infall/outflow/rotation) or turbulence injection (e.g. stellar winds).

Figure 7(b) shows the ratio of the $NH_3$(1,1), (4,4) and (5,5) linewidths for all cores detected from $NH_3$(1,1) to (5,5). While the $NH_3$(4,4)/$NH_3$(1,1) ratio is significantly greater than one, as discussed above, the $NH_3$(4,4)/$NH_3$(5,5) ratio shows these linewidths are much more similar so are likely to be tracing the same gas.

## 5.4 $NH_3$ (1,1) Quadrupole Hyperfine Anomalies

Some of the high resolution $NH_3$(1,1) core spectra show significant deviations from the predicted symmetric inner and outer satellite intensity ratios (see e.g. Rydbeck et al. 1977). Figure 8(a) shows an example spectrum overlayed with the predicted symmetric satellite ratios, which are labelled A to E in increasing velocity (i.e. A and E are the outer satellites, B and D are the inner satellites and C is the main line). The outer satellite lines, A and E, deviate significantly from the expected symmetry. Line profile asymmetries such as these are often seen toward star-forming cores (see Stutzki & Winnewisser 1985, and references therein) and only arise given very specific conditions. Qualitatively, the non-LTE conditions leading to the asymmetries can be understood by considering a large number of individual clumps within the beam, with much smaller linewidths ($\sim$0.3 kms$^{-1}$), and high density ($10^6$-$10^7$ cm$^{-3}$) whose combined spectra are averaged together into the observed spectra. In this scenario, the $NH_3$(1,1) level is still populated predominantly by rotational decay from the (J,K)=(2,1) transition, but the far-infrared photons released in the process undergo selective radiative-trapping. The individual clump linewidth of $\sim$0.3 kms$^{-1}$ means the transitions from the (J,K)=(2,1) level which populate the inner and main satellites of the $NH_3$(1,1) transition (B,C and D in Figure 8) overlap. Therefore, the escape probability of these photons (and hence the effective radiative rates populating the inner and main satellites) decreases, while the outer satellites (A and E in Figure 8) remain unaffected and so become over-populated. The over intensity of E with respect to A in the $NH_3$(1,1) spectrum then arises due to a combination of the available inversion transitions and their relative line intensities. If the individual clump line widths are greater than $\sim$0.3 kms$^{-1}$, the outer satellites will also overlap and the anomaly will be reduced. A lower limit to the number of clumps within the beam is given by the ratio of the observed linewidth to the linewidth of the individual clumps.

Park (2001) proposes a different scenario to produce $NH_3$(1,1) asymmetries. In this case, it is the velocity field from systematic motions within the core that is responsible for trapping the far-infrared photons. If the core is undergoing uniform/free-fall collapse, the velocity field is given by $V(r) \propto r$ and every ammonia molecule sees blue shifted emission from every other ammonia molecule. Therefore, photons emitted from lower energy rotational decays can be reabsorbed, but those from higher energy decays escape. The reverse is true for systematic outflow. Figures 8(b) and (c) show a schematic diagram of the predicted spectra from Park (2001) for infall and outflow respectively. The effect is the same, but reduced, if the velocity field is altered (e.g. for inside out collapse with $V(r) \propto r^{-0.5}$). This scenario requires the internal motions to be greater than 0.3 kms$^{-1}$ so the non-LTE asymmetries are not seen.

The hyperfine anomalies from the two scenarios can be differentiated by the ratios of their inner and outer satellites. If the anomalies are purely due to radiative trapping of non-LTE emission, E will be stronger A. However, systematic motions will affect both the outer *and inner* satellite lines with the result that both A and B are enhanced with respect to D and E for systematic infall and vice versa for systematic outflow (as illustrated schematically in Figure 8). This effect increases with $NH_3$ column density [$N_{NH_3}$] and hydrogen number density [$n(H_2)$].

To test for hyperfine anomalies we fit each of the five hyperfine components with an individual Gaussian to calculate their respective $T_A^*$. We then define three parameters, $\alpha$, $\beta$ and $\gamma$ from the ratio of the brightness temperatures of the individual components

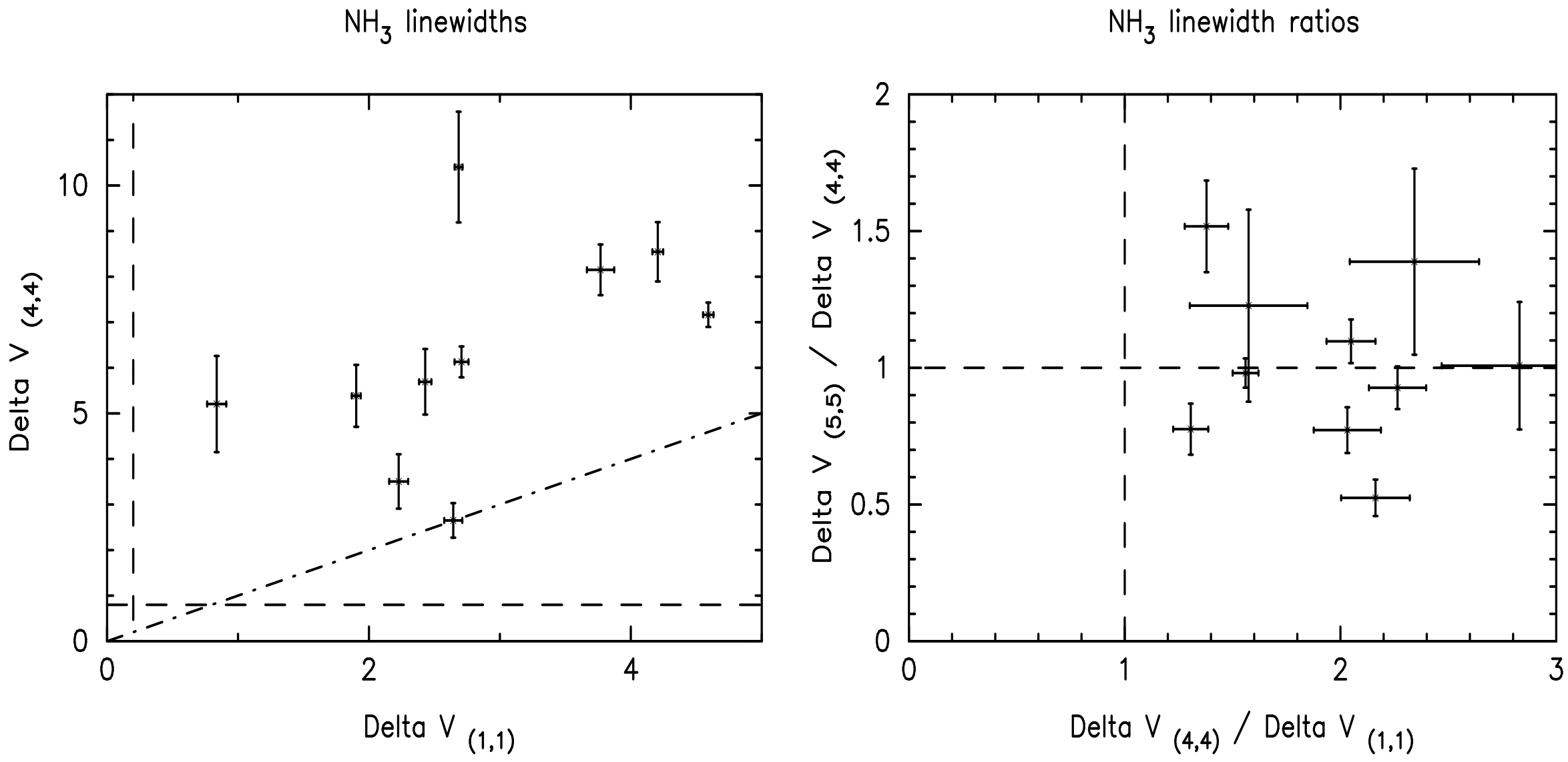

**Figure 7.** (Left) (a) $NH_3(4,4)$ vs $NH_3(1,1)$ linewidth for cores with detected $NH_3(4,4)$ emission. The vertical and horizontal dashed lines gives the spectral resolution of the $NH_3(1,1)$ and $NH_3(4,4)$ observations respectively and shows all the detections are well resolved. The dot-dashed line shows the expected relation if both transitions have the same linewidth. Clearly, the $NH_3(4,4)$ linewidths are significantly larger than those of the $NH_3(1,1)$ emission. (Right) (b) Ratio of the $NH_3(1,1)$, (4,4) and (5,5) linewidths for cores detected from $NH_3(1,1)$ to (5,5), with dashed lines to show a ratio equal to unity. While the $NH_3(4,4)/NH_3(1,1)$ ratio is significantly greater than one, the $NH_3(4,4)/NH_3(5,5)$ ratio shows these linewidths are much more similar.

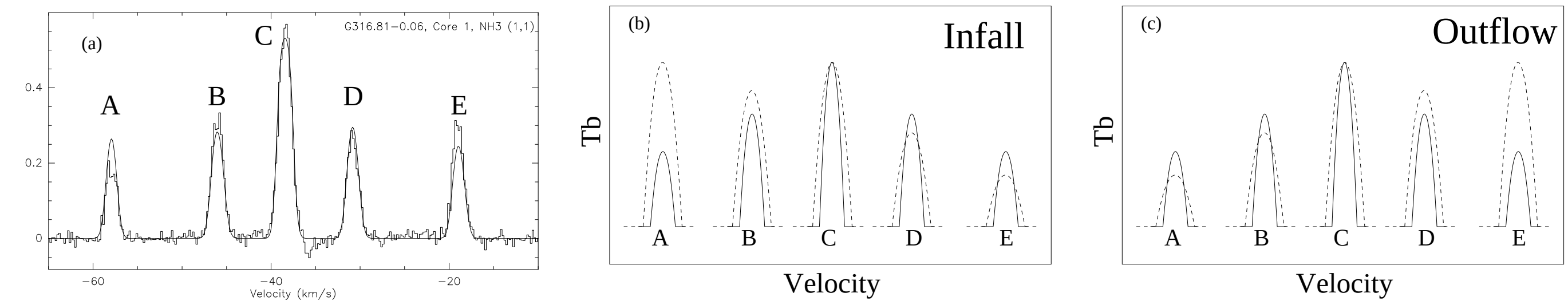

**Figure 8.** (a) Example of an observed $NH_3(1,1)$ spectrum with the five hyperfine components labelled A-E (as discussed in §5.4). The spectrum is overlayed with the fit to the hyperfine profile from the CLASS program, which illustrates that the predicted $NH_3(1,1)$ spectrum is symmetric about the main hyperfine line (C) i.e. the intenstities of A & E and B & D should be equal. However, the observed spectrum clearly deviates from this, with the intensity of E significantly greater than A. Modelling shows that this is due to non-LTE conditions within the emitting region (Stutzki & Winnewisser 1985). (b) and (c) show a schematic diagram of the theoretical, symmetric line intensities (solid line) and the results predicted by Park (2001) (dashed line) for cores undergoing infall (b) and outfbw (c). These anomalies due to systematic motions can be differentiated from those due to non-LTE conditions as they affect both the outer (A & E) and inner (B & D) satellites as opposed to just the outer satellites in the case of non-LTE conditions.

as follows,

$$\alpha = \frac{E}{A}$$

$$\beta = \frac{D}{B}$$

$$\gamma = \frac{D+E}{A+B}$$

In LTE, with symmetric profi les, $\alpha = \beta = \gamma = 1$ while for non-LTE conditions $\alpha > 1, \beta = 1$. For systematic infall we would expect both $\alpha, \beta < 1$ and outflow $\alpha, \beta > 1$. In the case of systematic motions, the larger the value $|\gamma|$, the more pronounced the systematic motion. The measured $\alpha$, $\beta$ and $\gamma$ values are shown in Table 8 and the uncertainties are from the formal errors to the individual Gaussian fi ts. The penultimate column shows the standard deviation in the measured linewidth of the fi ve lines and is used to fi nd bad fi ts in low signal to noise spectra. Cores with a standard deviation in the linewidth greater than 1 $\mathrm{kms}^{-1}$have been eliminated from further analysis. The fi nal column provides an interpretation of the remaining cores directly from their $\alpha$ and $\beta$ values. Figure 9 shows histogram plots of the derived $\alpha$, $\beta$ and $\gamma$ values and gives their mean and standard deviation. The $\alpha$ values range from 0.84 to 2.6 and are generally signifi cantly greater than 1, but the $\beta$ values, which range from 0.67 to 1.3, are much closer to 1. From Table 8 and Figure 9 it is clear that the majority of cores actually have $\alpha > 1$ but $\beta < 1$ – a result that neither of the above asymmetry scenarios predicts.

However, given the varying signal to noise of the spectra, a more rigourous criteria was used to test for asymmetries: as both anomalies described above have a more noticeable affect on the outer satellites, we defi ned a spectra as anomalous if $|\alpha - 1| >$ $3\Delta\alpha$ where $\Delta\alpha$ is the combined error in the Gaussian fits to A and E. With this criteria, 11 cores are found to have anomalous line ratios. The cores with the most pronounced asymmetries all belong to Group 1 (isolated $NH_3$).

Figure 10 shows the correlation plots between the $\alpha$, $\beta$ and $\gamma$ values with the mean error in the values either side of 1 shown as a dot-dashed line. They show that the cores with $\alpha$ significantly greater than 1, tend to have $\beta$ values less than 1. With values of $\beta$ closer to 1, the trend in $\gamma$ simply mirrors that in $\alpha$. As discussed above, this would suggest that the asymmetries are not due to systematic motions but rather non-LTE conditions within the cores.

### 5.5 Other signatures of systematic motion?

The characteristic $NH_3$(1,1) spectra of several cores show a double peaked profile, with a red or blue shifted absorption dip close in velocity to each of the five electric quadrupole hyperfine emission lines. Such profiles are often interpreted as signatures of systematic motions (e.g. Zhang & Ho 1997). Figure 11 shows an example of a red and blue shifted absorption spectra from G12.68-0.18 core 2 and G31.28+0.06 core 1. These appear very similar to the observed and theoretical spectra reported by Keto (1991) which were seen to indicate infall and outflow of molecular gas.

However, this effect is easily mimicked in interferometry data by spatial sidelobes of strong sources with emission at the (inverse) P-Cygni absorption frequency. With a relatively poor beam shape and complex source morphology (which are difficult to clean) higher fidelity images and complementary single dish spectra are required to determine if these signatures are real.

## 6 DISCUSSION

From the analysis in the previous sections, we have found the observed physical properties of the cores vary depending on their association with methanol maser and continuum emission. We now investigate what the physical conditions of the cores in the groups tell us about their evolutionary state.

As the core sizes within the groups are similar, the measured linewidths can be reliably used to indicate how quiescent the gas is, without worrying about its dependence on the core size (Larson 1981). It then becomes obvious that the isolated $NH_3$ cores with no methanol maser or continuum emission (Group 1) contain the most quiescent gas. However, from the linewidths alone it is not clear if these cores will eventually form stars or if they are transitory phenomenon. The fact that a large number of these Group 1 cores contain many dense sub-clumps (as evidenced by the $NH_3$(1,1) asymmetries) suggests the former is likely for at least some of the cores. Also, the linewidths are all greater than the (Larson 1981) relation for a given core size suggesting there may be some additional turbulence injection. The linewidth of the Group 2 ($NH_3$ + methanol maser emission) sources is larger than those of Group 1. In addition, the fact that only cores with methanol masers and/or 24 GHz continuum have $NH_3$(4,4) emission (which shows they are warmer than those without), suggests the cores in these Groups must be more evolved. The Group 3 ($NH_3$ + methanol maser + 24 GHz emission) sources, with much larger linewidths, stand out from the others as having the most turbulent gas. The detection of continuum emission suggests a massive star is already ionising the gas. It is therefore likely that the additional turbulence is generated from this internal powering source. Similarly, the Group 4 ($NH_3$ + 24 GHz emission) sources have significantly larger linewidths than those of Group 1. With the current observations, the properties of the continuum sources are not well enough constrained to further separate the evolutionary stages of the two groups. However, as all (with one possible exception) of the continuum sources only detected at 24 GHz are associated with dense molecular gas and masers, this would suggest they are younger, despite their seemingly small emission measures. Finally, the cores with only 8 + 24 GHz continuum, which make up Group 5, have no molecular material and are therefore sufficiently advanced for the UCHII region to have destroyed its natal material.

From this evidence, the cores in the different groups do appear to be at different evolutionary stages, going from most quiescent to most evolved roughly according to the group number. In subsequent work we will look to test this scenario through deriving further physical properties of the cores, including accurate kinetic temperatures from LVG modelling and higher resolution continuum observations.

## 7 CONCLUSIONS

We present data taken with the Australia Telescope Compact Array of para-ammonia [$NH_3$(1,1) $\rightarrow$ (5,5)] and 24 GHz continuum towards 21 southern Galactic hot molecular cores traced by 6.7 GHz methanol maser emission. We detect $NH_3$ toward every region. For each core we extract and fit characteristic $NH_3$ spectra and calculate the continuum properties. We find:

(i) $NH_3$(1,1) & (2,2) emission with similar morphology toward all 21 regions and 24 GHz continuum emission toward 12 of the regions, including 6 with no reported 8 GHz continuum counterparts.
(ii) 41 individual ammonia cores but around half of the regions only contain a single core.
(iii) $NH_3$(4,4) emission was detected in 13 of the $NH_3$(1,1) cores with coincident $NH_3$(5,5) emission toward 11 of these. This emission always peaks toward the $NH_3$(1,1) and the methanol maser position. Cores with $NH_3$(4,4) are warmer than those without rather than simply having more ammonia.
(iv) The $NH_3$ linewidth increases with transition energy. This additional broadening cannot simply be due to increased temperature so there may be contributions from either systematic motions (infall/outflow/rotation) or turbulence injection.
(v) The $NH_3$(1,1) spectra of several cores show a double peaked profile, with a red or blue shifted absorption dip close in velocity to each of the five electric quadrupole hyperfine emission lines. Higher fidelity images and complementary single dish spectra are required to determine if these spectra are signatures of systematic motions.
(vi) Several of the $NH_3$(1,1) spectra show large deviations from the predicted symmetric profile. We conclude this is due to non-LTE conditions possibly arising from a number of dense, unresolved clumps within the beam rather than systematic motions of gas in the cores.
(vii) 24 GHz continuum emission is observed toward 12 regions, including 6 with no reported 8 GHz emission. The 24 GHz only sources are all associated with $NH_3$ emission and some of the have flux densities which are consistent with optically thick free-free emission. We postulate these sources may be younger than those seen at 8+24 GHz, despite seemingly low emission measures, which could be an artifact of the large beam size. Alternatively, they may have been too extended and hence spatially filtered or resolved-out by the 8 GHz observations.

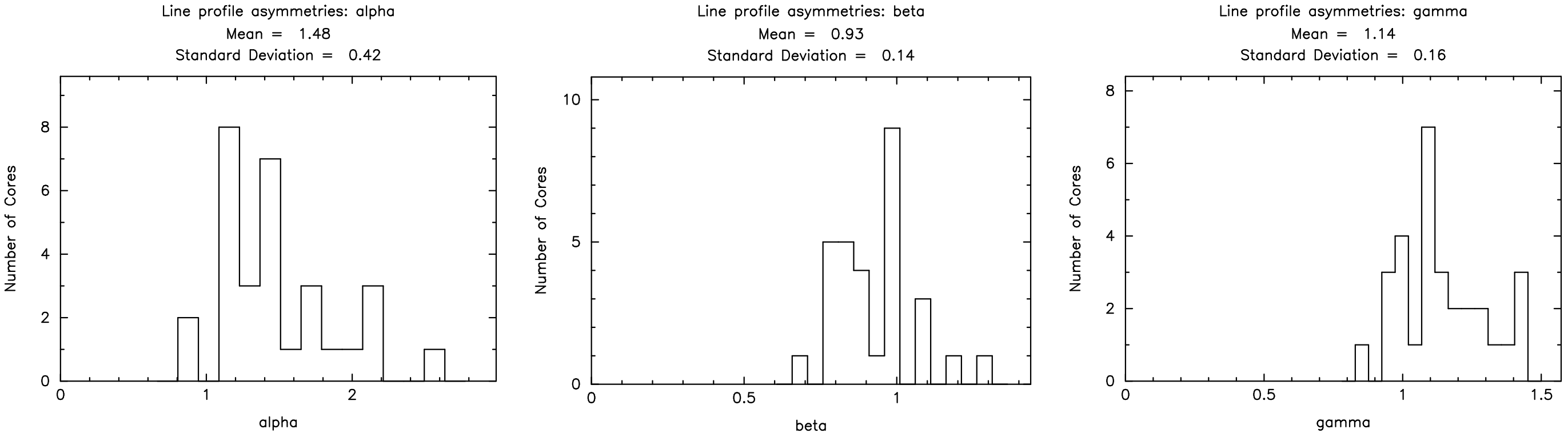

**Figure 9.** Histograms of the $NH_3$ (1,1) line profile assymetry parameters, $\alpha$, $\beta$ and $\gamma$ derived from the characteristic core spectra as outlined in § 5.4.

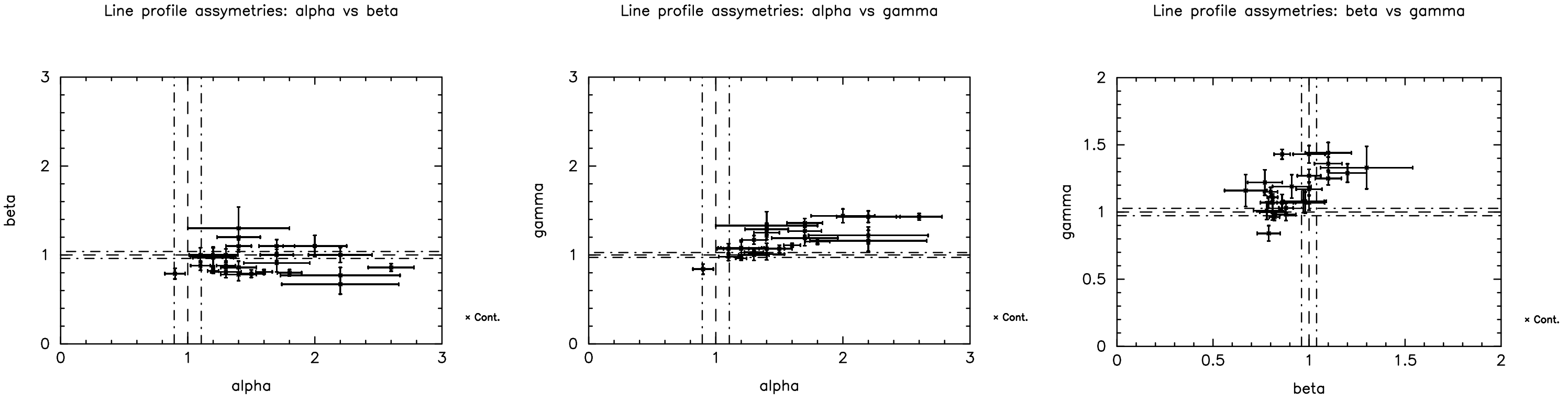

**Figure 10.** Correlation plots of the $NH_3$(1,1) line profile asymmetry parameters $\alpha$, $\beta$ and $\gamma$, for cores with asymmetric line profiles as outlined in § 5.4. The dot-dashed lines illustrate the mean error in the values either side of 1, which is shown separately as a dashed line.

(viii) The physical properties of the cores vary according to their association with methanol maser and continuum emission. From analysis of their physical conditions we postulate the cores may be at different evolutionary stages and propose a trend from most quiescent to most evolved as follows: isolated $NH_3$ cores (Group 1); $NH_3$ (1,1) cores associated with only methanol maser emission (Group 2); $NH_3$ (1,1) cores associated with 24 GHz continuum and methanol maser emission (Group 3); $NH_3$ (1,1) cores associated with 24 GHz continuum emission only (Group 4) and cores with 8 and 24 GHz continuum emission (Group 5).

## 8 ACKNOWLEDGEMENTS

SNL is supported by a scholarship from the School of Physics at UNSW. The Australia Telescope is funded by the Commonwealth of Australia for operation as a National Facility managed by CSIRO. This research has made use of NASA's Astrophysics Data System. We also thank the Australian Research Council for funding support from grant number DP0451893.

| Region | Core | $\alpha$ | $\beta$ | $\gamma$ | $\sigma(_{\Delta V})$ | Interpretation |
|---|---|---|---|---|---|---|
| G10.44-0.02 | 1 | 0.90±0.08 | 0.79±0.06 | 0.84±0.06 | 0.1 | – |
| G10.44-0.02 | 2 | 0.65±0.15 | 0.49±0.07 | 0.56±0.18 | 2.6 | – |
| G1.15-0.12 | 1 | 1.20±0.18 | 0.98±0.11 | 1.08±0.09 | 0.3 | non-LTE |
| G11.94-0.15 | 1 | 1.30±0.12 | 0.81±0.07 | 1.00±0.06 | 0.3 | non-LTE† |
| G11.94-0.15 | 2 | 1.40±0.14 | 0.78±0.07 | 1.01±0.07 | 0.2 | non-LTE† |
| G12.68-0.18 | 1 | 1.20±0.04 | 0.82±0.03 | 0.96±0.02 | 0.1 | non-LTE† |
| G12.68-0.18 | 2 | 2.60±0.18 | 0.86±0.04 | 1.43±0.04 | 0.1 | non-LTE† |
| G12.68-0.18 | 3 | 2.20±0.47 | 0.77±0.09 | 1.22±0.09 | 0.3 | non-LTE† |
| G12.68-0.18 | 4 | 1.30±0.28 | 2.10±0.34 | 1.79±0.17 | 2.7 | – |
| G19.47+0.17 | 1 | 1.40±0.15 | 2.00±1.20 | 1.62±0.39 | 29.8 | – |
| G19.47+0.17 | 2 | 0.95±0.28 | 1.20±0.18 | 1.05±0.17 | 1.2 | – |
| G23.44-0.18 | 1 | 1.30±0.07 | 0.88±0.04 | 1.03±0.03 | 0.3 | non-LTE† |
| G24.79+0.08 | 1 | 1.10±0.09 | 1.00±0.08 | 1.07±0.06 | 0.3 | NA |
| G24.79+0.08 | 2 | 1.20±0.14 | 0.98±0.10 | 1.07±0.08 | 0.2 | non-LTE |
| G24.85+0.09 | 1 | 1.80±0.10 | 0.80±0.04 | 1.15±0.03 | 0.2 | non-LTE† |
| G24.85+0.09 | 2 | 0.37±0.21 | 6.70±3.90 | 0.90±0.59 | 14.4 | – |
| G2.54+0.20 | 1 | 1.70±0.26 | 0.91±0.10 | 1.19±0.09 | 0.3 | non-LTE |
| G2.54+0.20 | 2 | 2.00±0.25 | 1.10±0.12 | 1.44±0.08 | 0.3 | non-LTE |
| G2.54+0.20 | 3 | 2.20±0.25 | 1.00±0.08 | 1.43±0.06 | 0.1 | non-LTE |
| G25.83-0.18 | 1 | 1.70±0.13 | 1.00±0.06 | 1.27±0.05 | 0.6 | non-LTE |
| G29.87-0.04 | 1 | 1.40±0.40 | 1.30±0.24 | 1.33±0.16 | 0.4 | NA |
| G29.96-0.02 | 1 | 1.40±0.10 | 1.10±0.07 | 1.25±0.05 | 0.3 | – |
| G29.98-0.04 | 1 | 1.70±0.14 | 1.10±0.07 | 1.36±0.05 | 0.3 | non-LTE |
| G30.79+0.20 | 1 | 1.30±0.09 | 1.00±0.07 | 1.17±0.05 | 0.2 | non-LTE |
| G31.28+0.06 | 1 | 1.20±0.16 | 0.97±0.11 | 1.08±0.09 | 0.1 | non-LTE |
| G316.81-0.06 | 1 | 1.60±0.07 | 0.81±0.03 | 1.11±0.02 | 0.1 | non-LTE† |
| G316.81-0.06 | 2 | 2.20±0.46 | 0.67±0.11 | 1.16±0.12 | 0.3 | non-LTE† |
| G323.74-0.26 | 1 | 1.40±0.17 | 1.20±0.10 | 1.29±0.07 | 0.5 | outfbw? |
| G323.74-0.26 | 2 | 1.10±0.08 | 0.88±0.05 | 0.98±0.04 | 0.3 | non-LTE† |
| G331.28-0.19 | 1 | 1.20±0.15 | 1.20±0.11 | 1.20±0.08 | 1.2 | – |
| G332.73-0.62 | 1 | 1.50±0.10 | 0.79±0.04 | 1.07±0.04 | 0.2 | non-LTE† |
| G332.73-0.62 | 2 | 1.40±0.14 | 0.86±0.07 | 1.07±0.06 | 0.1 | non-LTE† |
| G8.68-0.37 | 1 | 1.40±0.16 | 1.00±0.10 | 1.16±0.08 | 0.3 | non-LTE |
| G8.68-0.37 | 2 | 1.20±0.11 | 0.84±0.09 | 0.99±0.07 | 1.0 | – |
| G8.68-0.37 | 3 | 1.10±0.12 | 0.81±0.10 | 0.94±0.08 | 0.2 | non-LTE† |
| G8.68-0.37 | 4 | 0.84±0.07 | 1.00±0.07 | 0.93±0.05 | 0.3 | – |

**Table 8.** Brightness temperature ratios of the individual Gaussian profiles fit to the ammonia quadropole components (as outlined in §5.4). $\alpha$ and $\beta$ give the ratio of the outer and inner satellites respectively. $\gamma$ provides a measure of how skewed the profiles are from symmetric. $\sigma(_{\Delta V})$ gives the standard deviation in the line widths of the five components and is used to remove poor fits from further analysis. The final column lists the interpretation of the cores. 'NA' denotes the core is defined as statistically having no anomaly (i.e. being in LTE with no systematic motion). Cores with anomolous ratios are listed as either 'non-LTE' or having systematic 'infall' or 'outfbw'. A dash (–) denotes cases where it is either not possible to distinguish between the above interpretations, either due to multiple, blended velocity components or the standard deviation in the line width is greater than 1kms$^{-1}$. A, †, denotes that $\alpha > 1$ but $\beta < 1$, an anomaly which is not predicted either through non-LTE conditions or systematic motions.

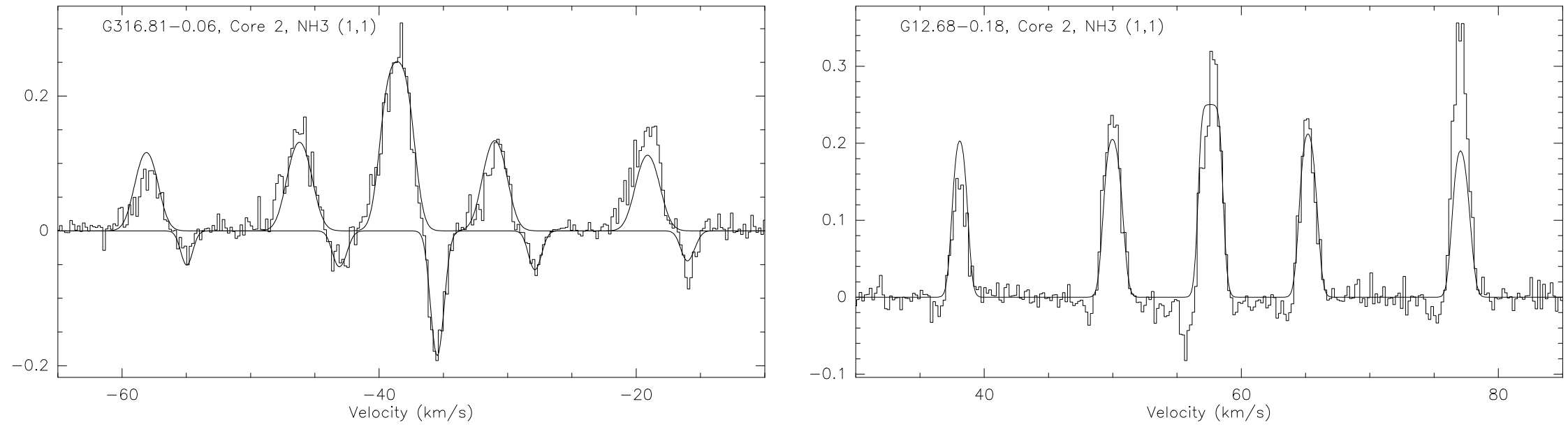

**Figure 11.** Example spectra of cores showing potential P-Cygni (Left) and inverse P-Cygni (Right) profiles which are often interpretted as signatures of infall and outfbw, respectively.

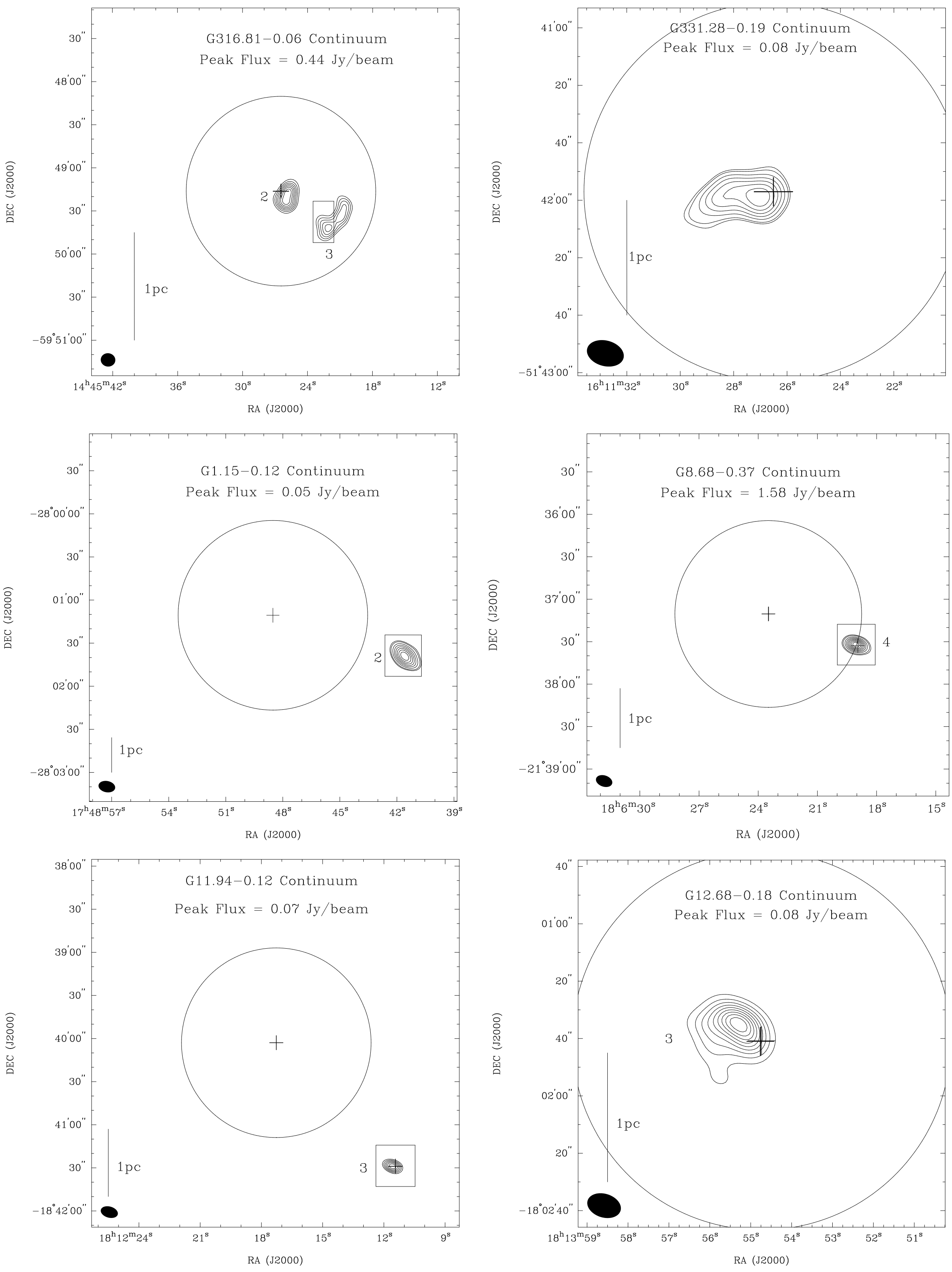

**Figure 12.** Maps of 24 GHz continuum emission with contour levels in steps of 10 per cent of the indicated peak value. The core number assigned to each emission region in the data tables is shown beside the relavent emission. Crosses and boxes show the position of methanol maser and 8 GHz emission respectively. The synthesised beam is shown in the bottom left corner of each image and the circles and lines give the primary beam size and linear scale respectively.

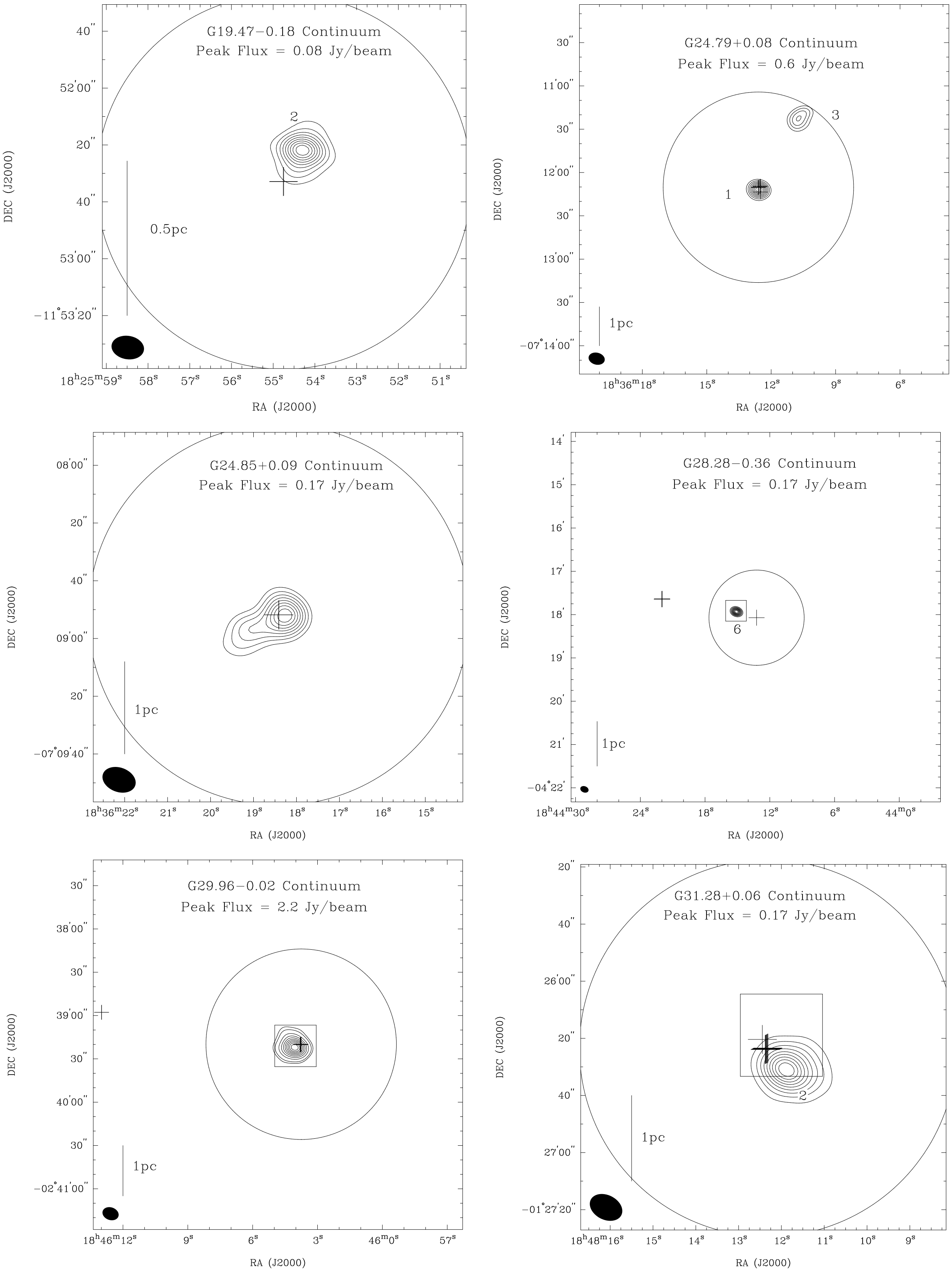

**Figure 12.** (Continued) Maps of 24 GHz continuum emission with contour levels in steps of 10 per cent of the indicated peak value. The core number assigned to each emission region in the data tables is shown beside the relavent emission. Crosses and boxes show the position of methanol maser and 8 GHz emission respectively. The synthesised beam is shown in the bottom left corner of each image and the circles and lines give the primary beam size and linear scale respectively.

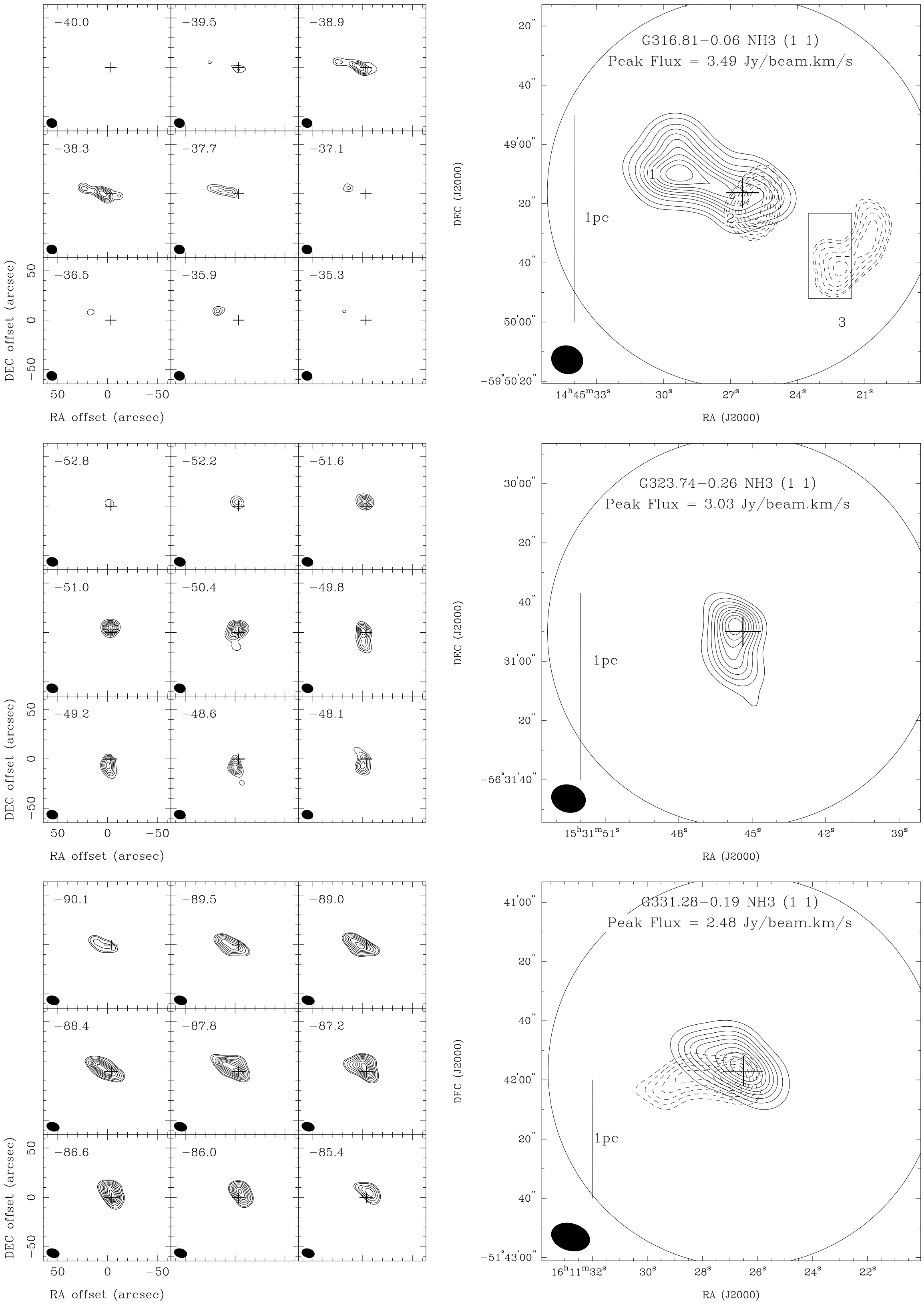

**Figure 13.** Example channel maps (left) and integrated intensity maps (right) of the $NH_3$ (1,1) emission. Contour levels are in steps of 10 per cent of the peak value indicated. G8.68-0.37 and G1.15-0.12 have an additional 5 per cent contour to highlight low intensity emission. The absolute position is given in the integrated intensity map and the offset from the central methanol maser position (in arcsec) is given for the channel maps. The $V_{LSR}$ (in $kms^{-1}$) of each channel map is given in the top left hand corner while the synthesised beam is shown in the bottom left corner of each image. Crosses and boxes show the position of the methanol maser and 8GHz continuum emission respectively. The circles and lines give the primary beam size and linear scale toward each region. 24GHz continuum emission is shown as dashed contours except the bold dashed contours in G8.68-0.37 and G12.68-0.18 which highlight regions of $NH_3$(1,1) absorption. Figures for the remaining regions are available in the published article or on request.

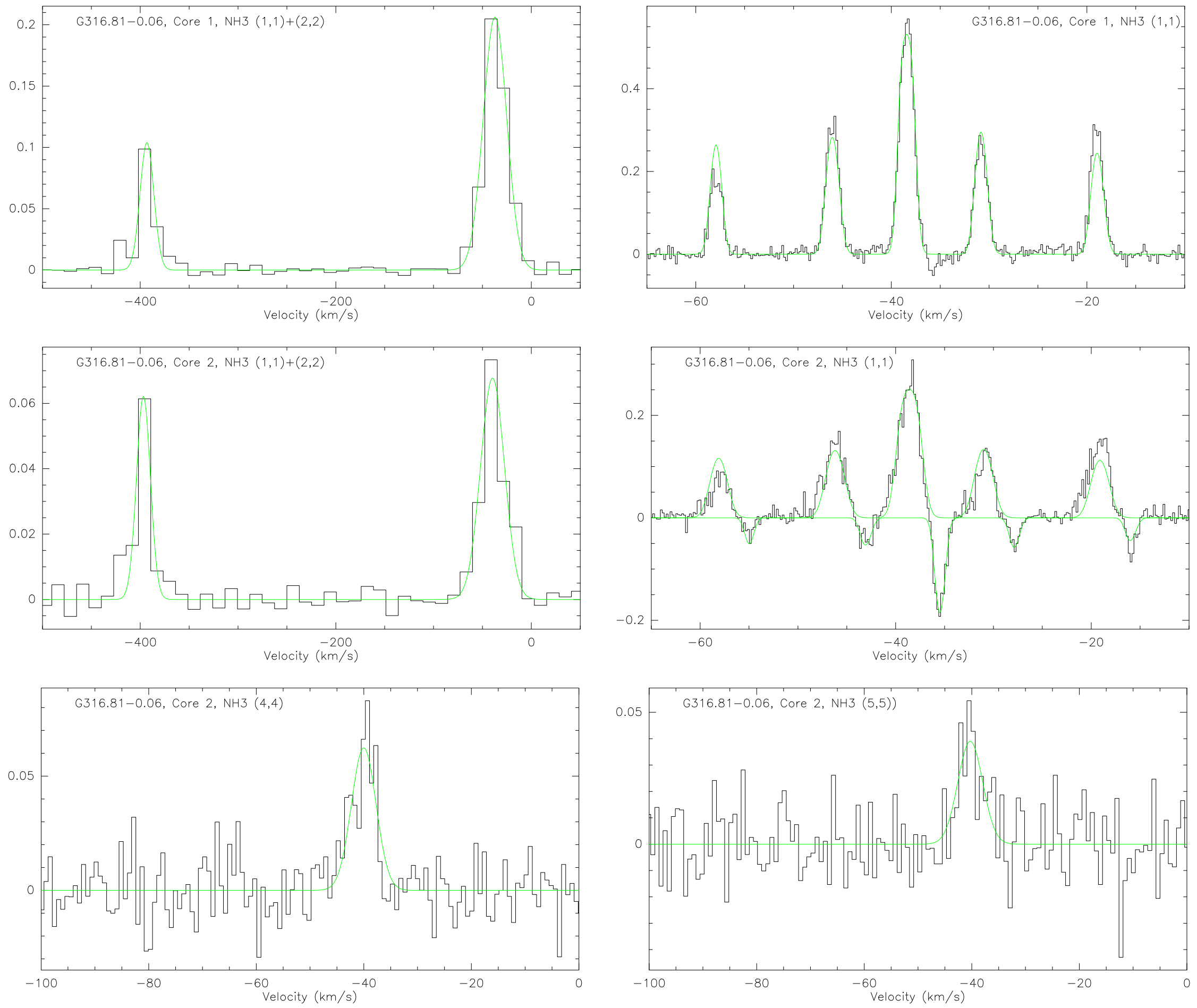

**Figure 14.** Example peak ammonia spectra from cores in source G316.81. The fit to the emission calculated in CLASS (as described in § 2) are overlayed on top of the spectra. The flux scale is in Jy. The remaining spectra are available in the published article or on request.